\documentclass[]{interact}

\usepackage{epstopdf}
\usepackage[caption=false]{subfig}

\usepackage{float}
\usepackage{multirow} 
\usepackage{threeparttable}
\usepackage{longtable}
\usepackage{etoolbox}
\usepackage{ragged2e}
\usepackage{makecell}
\usepackage{rotating}
\usepackage[style=apa, backend=biber]{biblatex}
\usepackage[utf8x]{inputenc}
\usepackage{amsmath}

\theoremstyle{plain}

\theoremstyle{definition}

\theoremstyle{remark}

\begin{document}

\articletype{ARTICLE TEMPLATE}

\title{Generative Artificial Intelligence-Guided User Studies: An Application for Air Taxi Services}

\author{
\name{Shengdi Xiao\textsuperscript{a}\thanks{CONTACT Shengdi Xiao, Jingjing Li. Author. Email: teki\_shengdixiao@digitalnature.slis.tsukuba.ac.jp, li@digitalnature.slis.tsukuba.ac.jp}, Jingjing Li\textsuperscript{bc}, Tatsuki Fushimi\textsuperscript{bc} and Yoichi Ochiai\textsuperscript{bc}}
\affil{\textsuperscript{a}Graduate School of Comprehensive Human, University of Tsukuba, 1-1-1 Tennodai, Tsukuba, Ibaraki, Japan;
\textsuperscript{b}Institute of Library, Information and Media Science, University of Tsukuba, Kasuga Campus Kasuga 1-2, Tsukuba, Ibaraki, Japan;
\textsuperscript{c}R\&D Center for Digital Nature, University of Tsukuba, Kasuga Campus Kasuga 1-2, Tsukuba, Ibaraki, Japan
}
}

\maketitle

\begin{abstract}

User studies are crucial for meeting user needs. In user studies, real experimental scenarios and participants are constructed and recruited. However, emerging and unfamiliar studies face limitations, including safety concerns and iterative efficiency. To address these challenges, this study utilizes a large language model (LLM) to create generative AI virtual scenarios for user experience. By recruiting real users to evaluate this experience, we can collect feedback that enables rapid iteration in the early design phase. The air taxi is particularly representative of these challenges and has been chosen as the case study for this research. The key contribution was designing a virtual ATJ using OpenAI's GPT-4 model and AI image and video generators. Based on the LLM-generated scripts, key visuals were created for the air taxi, and the ATJ was evaluated by 72 participants. Furthermore, the LLM demonstrated the ability to identify and suggest environments that significantly improve participants' attitudes toward air taxis. Education level and gender significantly influenced participants' attitudes and their satisfaction with the ATJ. Our study confirms the capability of generative AI to support user studies, providing a feasible approach and valuable insights for designing air taxi user experiences in the early design phase.

\end{abstract}

\begin{keywords}
Generative AI, Large Language Model (LLM), User Studies, Virtual Experimental Scenarios, User Experience (UX), Air Taxi
\end{keywords}

\section{Introduction}

\begin{figure}[H]
\centering
\includegraphics[width=\linewidth]{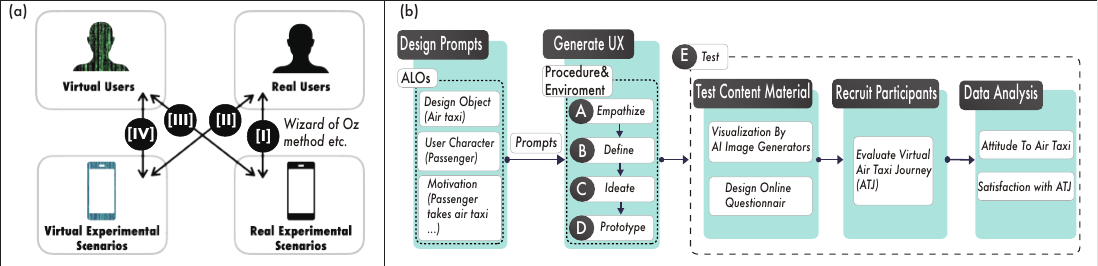}
\caption{The summary of this study. (a) The main contribution of this study. [I] means a user study conducted by real experimental scenarios and real users, [II] means a user study conducted by virtual experimental scenarios and real users, [III] means a user study conducted by real experimental scenarios and virtual users. [IV] means a user study conducted by virtual experimental scenarios and virtual users. This study focus on a user study with virtual experimental scenarios (air taxi) and real users. (b) The research structure of this study.} \label{fig:1-1}
\end{figure}

User studies, which are experiments involving human participants, are informal evaluations of user interfaces (UIs) or interactions \parencite{arbelaez2016improving, lyly2019mock, venkatesh2012consumer}. It is an indispensable research method in Human-Computer Interaction (HCI). Researchers use it to gather insights and data about user interactions with a particular product, system, or environment to understand their behaviors, preferences, and needs, and to ensure that the technology meets user expectations while providing a satisfying interaction. To achieve a comprehensive understanding of target users, several conditions are necessary. These include interviewing target users, setting up the experimental environment, building prototypes, recruiting participants to evaluate the system, and having supervisors observe the process to collect accurate data \parencite{kuniavsky2003observing}. However, physical prototypes are needed for user studies in the field of innovations and technologies, such as the transportation service \parencite{astfalk2021effect, camburn2017design, Stowe2008InvestigatingTR}. Enabling real users engagement in real experimental scenarios for experience and evaluation is ideal for developing transportation applications, but the instability of new transportation technology in pre-tests can pose safety issues. As shown in Figure~\ref{fig:1-1} (a), it highlights four key factors: the virtual and real aspects, and the user and the experimental scenario. Previous user studies in transportation, as depicted by the [I] and [II] in Figure~\ref{fig:1-1} (a), have utilized real and virtual experimental scenario, respectively, but both approaches face challenges related to safety and iterative efficiency. Regarding the [I], the following relevant studies currently exist. For example, studies involving real air taxis have demonstrated the flight of a real air taxi to assess public acceptance \parencite{unknown}. For autonomous taxis, Wizard of Oz methodologies with real vehicles have been employed to explore user experience design \parencite{taxi234, meurer2020wizard}. 

To address these concerns, many researchers have turned to virtual experimental scenarios [II]. For instance, the study has explored the interactions between self-driving vehicles and cyclists in immersive virtual reality (VR) \parencite{10.1145/3313831.3376884}. VR has been used to design services related to air taxis \parencite{kim2024designing,birrell2022urban}. Additionally, other studies have provided participants with images of flight scenarios to gauge perceptions of air taxis \parencite{arbelaez2016improving,ward2021safety}. Despite these efforts, creating virtual experimental scenarios remains time-consuming and resource-intensive, as it requires developing comprehensive UIs and cockpit designs. This represents only a fraction of the early design phase, necessitating iterative updates based on user feedback. While iterative refinement is crucial, it imposes significant strain on the resources and efforts of developers, thereby hindering the rapid advancement of the technologies. Furthermore, for iterative processes in the early design stages, it increases the likelihood of users encountering safety issues during the testing process due to the immaturity of the technology. Yet, there is a dearth of studies addressing these issues.

We propose using the generative AI, specifically LLM and AI image and video generators, to rapidly and efficiently design the virtual experimental scenarios, enabling quicker and more effective user study iterations in the early design phases [II]. With the emergence of LLM, which is typically trained on enormous internet datasets, such as Common Crawl \parencite{raffel2020exploring}, there have been some new advances in research on HCI. To explore the capability of generative AI in creating virtual experimental scenarios thereby optimizing the user study process, we recruited real users to experience and evaluate the virtual experimental scenarios designed with generative AI in this study. Obtaining feedback from real users that closely mirrors real-world responses evaluating virtual experimental scenarios is anticipated to enhance the design of real experimental scenarios. Therefore, this study follows the following five steps, which are Design Prompts, Generate UX, Test Content material, Recruit participants, and Data analysis (see Figure~\ref{fig:1-1} (b) ). For demonstration purposes, we selected air taxis as a case study due to the unique challenges they present. First, we designed prompts utilizing OpenAI’s GPT-4\footnote{https://openai.com/research/gpt-4} model to identify and suggest the virtual experimental scenarios including the user experience from calling to dropping off the air taxi, mockup and storyboard. Key supplementary graphics and videos of the Air Taxi Journey (ATJ) were generated by Midjourney\footnote{https://www.midjourney.com/} and Runway\footnote{https://runwayml.com/} based on scripts generated by GPT-4. Second, we recruited 72 participants to experience and evaluate the ATJ.

This study contributes to the discourse on the application of generative AI to conduct virtual user studies. Furthermore, by simulating the user experience, we aim to bridge the gap between technological capability and user acceptance. Consequently, these outcomes could serve as invaluable insights for the designer and researchers in the earlier stages of their iterative design processes. Future research may explore the potential of using LLM-generated user responses for evaluating both real and virtual experimental scenarios, as depicted in Figure~\ref{fig:1-1} (a) [III] and [IV]. It is expected to address challenges in HCI research, including low reproducibility, high barriers for ethical review, and the necessity of controlled test environments. The rest of this paper is organized as follows: Section 2 presents a review of the relevant literature on user study, the role of LLM and AI image and video generators, case study towards air taxis. Section 3 describes the methodology, including experimental design principle and procedure, and user test. Section 4 presents the results and the data analysis. Section 5 discusses the implications of our findings, potential limitations, and future research directions. Finally, Section 6 concludes the paper and highlights its main contributions.

\section{Related Work}
\subsection{Current User Study Methods}\label{class}

User studies are a fundamental component of HCI research, essential for understanding user behaviors, preferences, and interactions with various technologies \parencite{lazar2017research}. The studies are employed in a variety of methods to understand and collect user feedback. Traditional methods such as survey, interview \parencite{totawar2016research, dumas1999practical}, and case studies are commonly employed in user study. Although effective, these methods can be time-consuming and resource-intensive \parencite{10.1145/1066348.1066325}. Moreover, collecting verbal data, as noted by Hämäläinen et al., can be both slow and costly \parencite{10.1145/3544548.3580688}. Prototypes are created to support these experimental scenarios, allowing participants to experience and evaluate them \parencite{sun2013comparison}. They are often used to gather user feedback before technologies are market-ready, but developing physical prototypes is resource-intensive and typically occurs later in the development process. These impact the iterative efficiency. Additionally, it is fundamental for designers to prioritize safety while understanding the cognitive and emotional characteristics of users \parencite{gould1985designing}. The immaturity of certain technologies in user studies often leads to increased safety concerns during the early design phase. These studies require more efficient iterations to enhance the technology safely.

To mitigate the iterative efficiency and safety issues commonly encountered in user studies, this study leverages LLM and AI image and video generators to generate virtual experimental scenarios. These scenarios are tailored to user needs, including UIs, storyboards, and experimental contexts. This approach aims to streamline the iterative process by providing a safe and efficient environment for user studies, thereby enhancing the overall design and usability of the technology.

\subsection{The Role of LLM and AI Image and Video Generators}\label{class}

With the emergence of generative AI, the development of HCI has changed. LLM, such as GPT-4, can simulate human comprehension and predict sequences in various text formats, providing a reliable foundation for the HCI experience \parencite{10.1145/3544548.3580688, brown2020language}. Google launched the latest multimodality AI-Gemini\footnote{https://gemini.google.com/?hl=ja}, which can perform excellently in image, audio, video, and text understanding \parencite{team2023gemini}. This improves the capabilities of LLM in the “Video question answering” feature. Abstract language objects (ALOs) were proposed by Ochiai et al. \parencite{ochiai2023towards} in the field of LLM. ALOs are prompting technologies for use in GPT-4, programming, and ubiquitous computing scenarios. It is possible to use ALOs to establish connections between the linguistic objects and classes in GPT-4 and their counterparts in high-level programming languages. 

The advanced text generation and creative features of ChatGPT render it more adept than Gemini at crafting concise summaries and facilitating natural language interactions with users \parencite{rane2024gemini}. GPT-4 and Gemini were used to simulate the responses of the participants in the user test. The ability of the Gemini to process and interpret video data allows for the replication of human participants’ viewing conditions, thus maintaining controlled consistency in experimental settings. However, the simulation experiment demonstrated the comprehensiveness of the LLM. In the medical field, a study by Carlà et al. demonstrated that Google Gemini and GPT-4, through consistent evaluation of vitreoretinal patients' records, achieved substantial agreement with expert surgeons \parencite{carla2024exploring}. Therefore, the variation in simulation outcomes between image and video analyses may also expand the application of these technologies to the transportation sector, beyond their current use in the medical field . The generated AI possesses the power to create and its media generation process functions like the human creation process. Some popular tools, such as Midjourney and Runway, have accelerated this process, in which computers can synthesize images based on textual descriptions \parencite{manovich2023ai}. Accordingly, the images and videos required for the evaluation experiments were generated using the Midjourney and Runway software.

\subsection{Case Study: Air Taxis}\label{class}

\subsubsection{The Current Development of Air Taxis}\label{class}
Air taxi, also known as Urban Air Mobility (UAM), enabled by vertical take-off and landing (VTOL) capabilities, has emerged as a potential mode of transportation to address traffic gridlocks within cities \parencite{rajendran2020air}. It offers the potential for faster travel times, enhanced convenience, reduced emissions through power and energy requirements that are minimized using low-disk-loading rotors \parencite{johnson2018concept}, and efficient utilization of three-dimensional airspace. Thus, air taxi services attract concern from diverse stakeholders, such as industry, government, and academia \parencite{binder2018if, rajendran2020air}. The United States decided to build infrastructure for air taxi services \parencite{newspaper2021} and the Brazilian airline Gol plans to launch a fleet of electric air taxis that are cheap and have zero emissions as Uber \parencite{newspaper2021-2}. Investors pledged a record USD 4.3bn for air taxi start-ups \parencite{newspaper2021-1}. The air taxi design must incorporate safety features and recommendations mandated by local government regulations \parencite{rajendran2019insights}. Recent studies have increasingly focused on developing new designs for air taxis that adhere to safety regulations set forth by governmental authorities \parencite{piwek2016small}. The academic community is currently enhancing the development of air taxis through numerous dedicated studies.  

\subsubsection{User Study of Air Taxis}\label{class}

Several studies have used diverse methods to investigate aspects related to air taxi design, user acceptance, and public perceptions of developing air taxis. Through an online survey, this study collected 692 responses on public perceptions of flying cars and taxis, providing insights into the attitudes and acceptance of potential users \parencite{Surveydata}. Ahmed et al. conducted an online survey to explore the extent to which people are willing to rent and pay for flying taxis and share flying car services \parencite{ahmed2021exploratory}. In another study employing online focus groups, participants universally identified cabin design, noise levels, window configurations, and room temperature as significant factors \parencite{stolz2021user}. In addition, a study by Schiffer underscore the fact that a physical demonstration of air taxi is not necessary to gain a comprehensive picture of people’s attitudes before full development \parencite{astfalk2021effect}. Consequently, we provided the participants with the virtual experimental scenarios, referred to as ATJ, to support the air taxi user studies in preparation for the design of real experimental scenarios.

Before introducing new technology into the market, fewer resources, funding, effort, and time are required to drive an iterative design process, which is critical, particularly in the early design phase. Although, as previously stated, air taxi developers have already attempted various approaches to advance the field, current research has not yet effectively addressed the rapid iteration required for early-stage research. Furthermore, the public has safety concerns about air taxi \parencite{al2020factors, goyal2018urban}, which would impact the users' evaluation and feedback on air taxi. Therefore, air taxis are particularly representative of these challenges and has been chosen as the case study for this research. 

\section{Methodology}

\subsection{Experiment Design Principle and Procedure}

In this experiment, a design-thinking process was applied to explore user needs. The design-thinking process proposed by the d.school at Stanford University (2010) is a five-step iterative process that includes empathize, define, ideate, prototype, and test \parencite{carlgren2016framing}. Design Thinking has been integrated into the collective consciousness of design researchers \parencite{rowe1991design} and is an approach to thinking and logic applied throughout the design process. It is a human-centered design approach for deriving innovative concepts based on an in-depth understanding of human needs \parencite{article1}. This can help designers identify research problems and develop more effective solutions \parencite{designs4010004}. The field of design research has not been inclined to simplify the research object but rather to value multiple perspectives and a rich picture \parencite{dorst2011core}. Furthermore, as mentioned in the previous section, it is critical in the early stages of design to continually improve the overall design through rapid iterations to develop new technologies. Hence, we adopted the design-thinking process to design the experimental procedure, and GPT-4-driven prompts (see Appendix B) conducted this five-step iterative process for the air taxi user study. Each step of the experimental procedure is described as follows.  

Based on the design thinking design process, this study gives the way for how to build the virtual experimental scenarios. ALOs were set as design object (product or service), user character (end-users), and purpose (user engagement with the product or service). This linguistic object serves as the foundation for the subsequent steps in our approach, including object interaction, simulation, and integration with high-level programming languages. Then we employed GPT-4-driven prompts to guide us through the empathize, define, ideate, and prototype stages of the five-step iterative design thinking process. The virtual user characteristics appropriately address potential issues in the system to enhance efficiency since real participants are not involved at this stage. AI image and video generators were then utilized to visualize the experimental scenarios in the user experience to conduct user tests. Finally, real participants experienced and evaluated these scenarios, providing their feedback.

\begin{figure}[H]
\centering
\includegraphics[width=\linewidth]{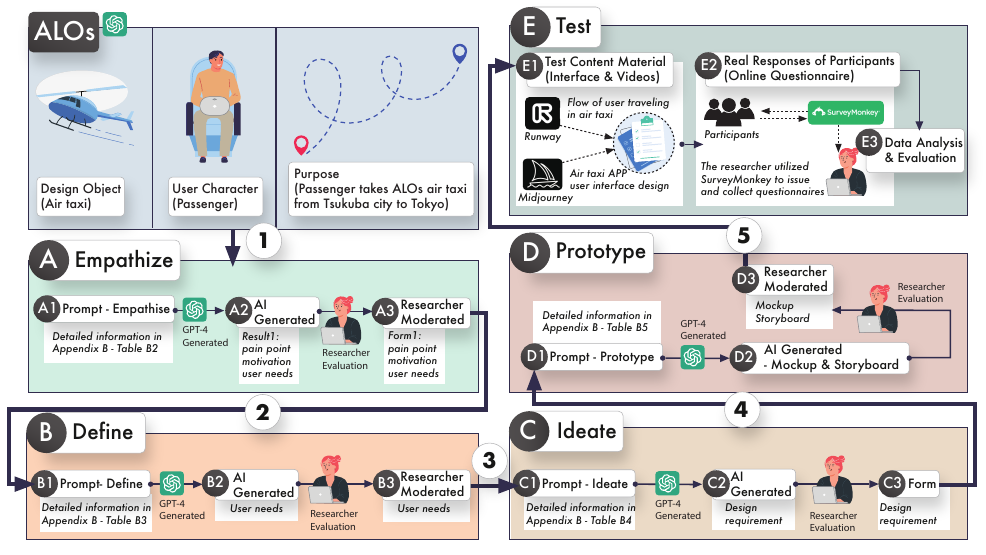}
\caption{Procedure of how we conducted the user study using LLM and AI image and video generators. A, B, C, D, and E represent the five-step iterative design thinking process respectively, including empathize, define, ideate, prototype, and test. A1, B1, C1, and D1 represent the prompts we designed for each step. A2, B2, C2, and D2 represent the outputs from GPT-4. A3, B3, C3, and D3 represent the inputs into GPT-4. In the E-Test step, E1 is the necessary test content material for the user test. E2 is the real responses from participants’ evaluation towards ATJ. E3 represents the data analysis of towards evaluated responses step.} \label{fig:2}
\end{figure}

The detailed method of application to case air taxi is described. As shown in Figure~\ref{fig:2}, using OpenAI's GPT-4 to define the ALOs as the system definition of the API by “Prompt-ALOs” (see Appendix C - Table~\ref{tab:freq-31}) and input “ALOs (air taxi), ALOs (passenger), and ALOs (passenger takes the air taxi from Tsukuba to Tokyo)” firstly. Secondly, user pain points, motivations, and user needs for air taxis (see Appendix C - Table~\ref{tab:1}) were analyzed through “Prompt-Empathize” (see Appendix B - Table~\ref{tab:freq-1}). Subsequently, the appropriate result selected by the researcher as the result of empathizing was inputted into “Prompt-Define” The user needs (see Appendix C - Table~\ref{tab:2}) were then generated from the “Prompt-Define” (see Appendix B - Table~\ref{tab:freq-2}). Then, the appropriate results selected by the researcher and input them into “Prompt-Ideate” (see Appendix B - Table~\ref{tab:freq-3}) to generate the design requirement (see Appendix C - Table~\ref{tab:3}). Next, the appropriate result selected by the researcher as the result of ideation and input into “Prompt-Prototype” (see Appendix B - Table~\ref{tab:freq-4}) to generate the corresponding mockup (see Appendix C - Table~\ref{tab:4}) and storyboard (see Appendix C - Table~\ref{tab:5}) for the user test. The appropriate results selected by the researcher as the prototype and utilized in the test. These are the first 4 processes of design thinking in GPT-4 using our designed prompts.

For the “Test” step (Step E), the necessary material about the mockup and storyboard was visualized by Midjourney and Runway for inputting into the evaluated questionnaire as the evaluated result of “Prototype” was textual (Step E1). 72 participants were recruited to evaluate the simulated journey online(Step E2). The data analysis (Step E3) was conducted in Section 4. 

Through data analysis (Step E3), the reliability of our designed methodology relied on the LLM and AI image and video generators was confirmed. We propose this generative AI-guided method, inclusive of the devised prompts, as a reference tool for conducting user studies in the early design phrase. Furthermore, it is envisaged to facilitate workshops in UX design across various fields. The efficacy and broader applicability of generative AI-guided approaches invite future validation and utilization by researchers. 

\subsubsection{Prototype Design}

The textual prototype of the mockup and storyboard toward the air taxi was generated in Steps A to D. The suitable output contents we selected are shown in Appendix C - Table~\ref{tab:freq-5} and Table~\ref{tab:freq-6}. We explained in detail how the visualization process works for user tests in Section $3.2.1$.

\begin{figure}[H]
  \centering
  \includegraphics[width=\linewidth]{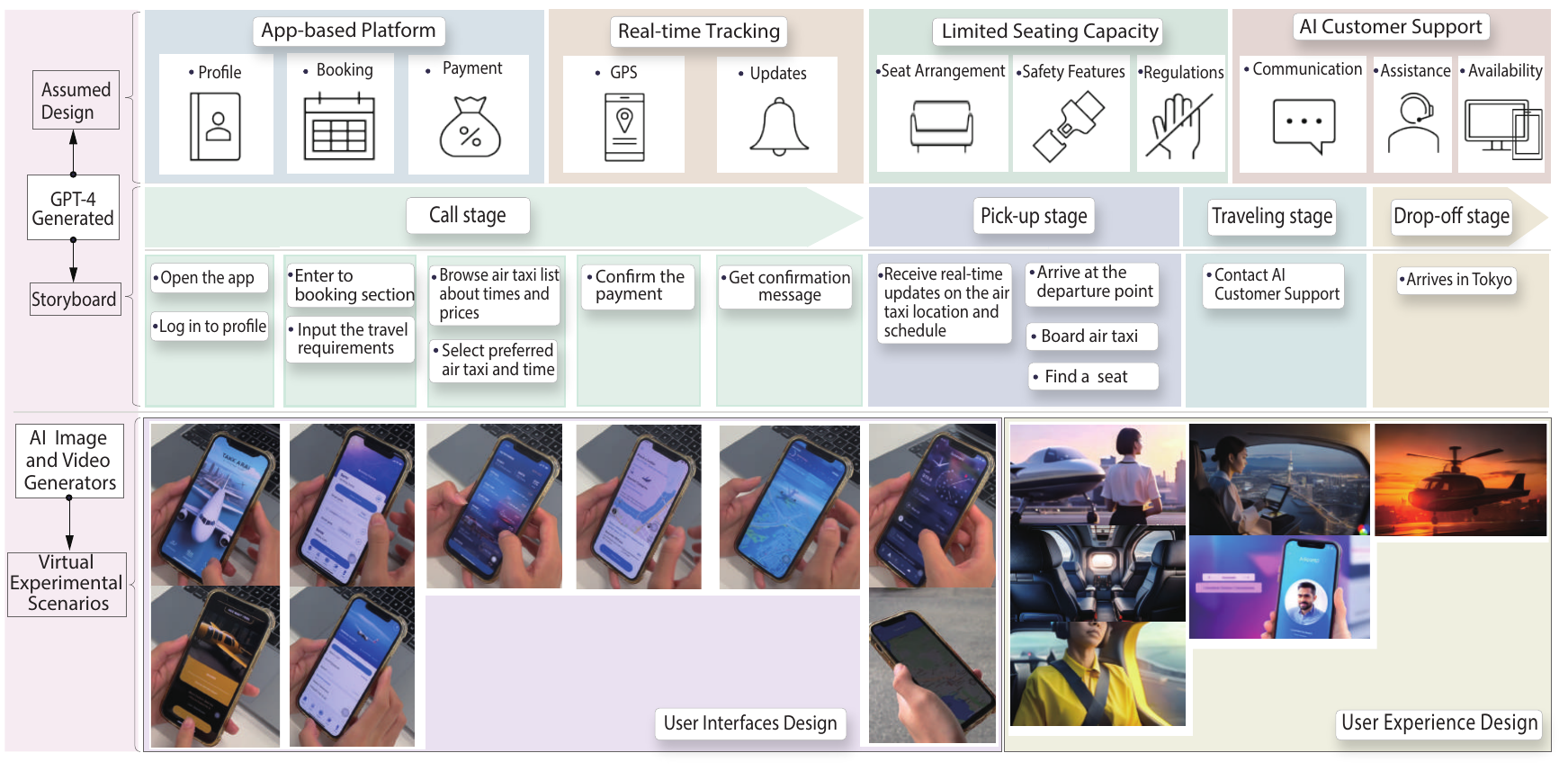}
  \caption{A customer journey map. It is mainly summarized based on the selected mockup (in Appendix C - Table~\ref{tab:freq-5}) and storyboard (in Appendix C - Table~\ref{tab:freq-6})}
  \label{fig:8}
\end{figure}

Customer journey map is a prevalent strategic tool that promotes an understanding of customer experiences for both academics and practitioners \parencite{rosenbaum2017create}. Therefore, the air taxi customer journey map depicted in Figure~\ref{fig:8} illustrates the interaction between passengers and air taxis during the trip based on the above prototype of the mockup and storyboard. The journey begins when the call air taxi stage drops off. During the call stage, the app-based platform provides a profile, booking, and payment section. Passengers can log in to their personal profiles, booking trips, confirming payments, and receiving messages. On the pick-up day, real-time tracking, including GPS functions and updates, can remind passengers of the air taxi location and schedule. Subsequently, limited-seating-capacity air taxis, including seating arrangements, safety features, and regulations, are presented to passengers. During the traveling stage, AI customer support was available to support passengers. 

Our design was intended to cater to the basic needs of air taxi developers in the initial stages of design, and we plan to evaluate the specific interactions between passengers and air taxis in a more refined manner in future studies. The customer journey map shown in Figure~\ref{fig:8} illustrates various elementary interactions between passengers and air taxis, demonstrating the potential of our designed prompt to provide valuable information to air taxi developers regarding different interactions. 

\subsection{User Test (Step E)}

\subsubsection{Questionnaire Design}

In Section $3.1$, we describe the process of conducting a user study of air taxis using GPT-4, which includes empathize, define, ideate, and prototype. Following the above process, GPT-4 provided suggestions for a literal mockup and storyboard for air taxis. This shows that the LLM can be used to conduct user studies on air taxis. However, both the air taxi developer and LLM designed the mockup and storyboard of air taxis, which needed to be evaluated and tested by user tests. Therefore, we visualized them using AI image and video generators and photo equipment placed on a questionnaire for real-life user tests. 

Midjourney was used to visualize the mockup generated by GPT-4 (in Appendix C - Table~\ref{tab:freq-5}) in UIs as material. The iPhone was then used to record user interactions with these UIs of the storyboard (in Appendix C - Table~\ref{tab:freq-6}). The remaining were visualized using Runway (in Appendix C - Table~\ref{tab:freq-6}). Finally, a visualized storyboard was generated as the virtual experimental scenarios for the user test (see Figure~\ref{fig:8}). We referred to the entire experience of taking air taxis as “air taxi journey (ATJ)” and used it in the rest of the paper.

The questionnaire was designed to evaluate users using a 5-point Likert scale. It is divided into three modules: personal information, ATJ evaluation, and attitudes toward air taxis. Personal information included participants’ nationality, age, gender, education level, employment status, personal interests, and transport preferences. This module was used to accurately simulate real participants, enabling the execution of virtual participant evaluations in the LLM. The evaluation of the ATJ module is a visualization of taking air taxis, from calling to dropping them off. Attitudes toward the air taxi model explore the influence of ATJ on people’s attitudes. Both models were designed to demonstrate the validity of the LLM in conducting a user study of air taxis. The specific content of the questionnaire is presented in Appendix A.

\subsubsection{Participants}

A priori power analysis was conducted to detect the minimum sample size required for testing our study by applying G$\ast$ Power version $3.1.9.6$ \parencite{faul2007g}. The result suggested that a sample size of 54 was required to achieve 95\% power to detect a medium effect at a significance level of $\alpha=0.05$. A total of 72 participants were recruited for the study (40 males and 32 females). As this experiment was conducted to ensure sample diversity, 72 participants were recruited for user test. As shown in Table~\ref{tab:freq-40}, the sample was divided into four groups based on gender, educational level, age range, and employment status. Education level was categorized into high school or lower, college, Bachelor’s degree, Master’s degree, Ph.d.~, or higher. Their ages ranged from 18 to 59 years, and they were classified into two groups: 18$\sim$40 years and \textgreater 40 years. Employment status was categorized as unemployed or employed (see Table~\ref{tab:freq-40}).

\begin{table}[H]
\centering
\caption{Distribution of the sample according to demographic characteristics}
\label{tab:freq-40}
\begin{tabular}
{p{3.2cm}<{\centering} p{3.2cm}<{\centering} p{2.5cm}<{\centering} p{3cm}<{\centering}}\\ \hline
Variable (N=72)& & Frequency& Precentage \\ 
\hline
Gender & & &\\ 
 &Male&40 & $55.6$\\ 
 &Female& 32& $44.5$ \\ 
\hline
Education level & & &\\
 & high school or less&9 &$12.5$\\
  & some college&15 &$20.8$\\
   & Bachelor's degree&28 &$38.9$\\
    & Master's degree&14 &$19.4$\\
     & Ph.d. or higher& 6&$8.3$\\\hline
     Age Range & & &\\
       & 18$\sim$40& 51&$70.8$\\
        & 40+& 21&$29.2$\\\hline
Employed status & & &\\
 & Unemployed& 9&$12.5$\\
  & Employed& 63&$87.5$\\\hline
\end{tabular}

\end{table}

\subsubsection{Procedure: Real responses from Participants (Step E2)}

As described in Section $3.2.2$, a total of 72 participants were recruited, which is in line with the number of participants required in HCI studies. Each participant was asked to answer questions about their attitude towards air taxis before and after viewing the video ATJ on a 5-Point Likert scale ($1=$“Not at all”, $5=$“Very much”). In addition, after viewing each scene, participants were also asked to rate each scene on a 5-point Likert scale ($1=$“Not at all”, $5=$“Very much”). Please check the specific content of the questionnaire in Appendix A. Finally, we analyzed the data on the responses received from the participants, which are presented in Section 4.

\section{Results and Data Analysis}

In this section, we summarize the participants’ attitudes toward air taxis, their overall satisfaction with the flight simulation, and the differences between their real and virtual responses. 

\subsection{The Attitude to Taking Air Taxi}

We examined the quality of the ATJ based on respondents’ attitudes toward taking an air taxi before and after watching the ATJ. As shown in Table~\ref{tab:freq-7}, the data indicate an improvement in people’s attitudes toward taking an air taxi after watching the ATJ, with the mean value of attitudes toward taking it increasing from $3.46$ to $3.89$. In addition, a paired samples t-test revealed that there was a statistically significant positive difference in participants’ attitudes to taking an air taxi after viewing ATJ ($t=-3.326$, $p=0.001$*). This implies that the generative AI can effectively identify and suggest air taxi environments, which improves the participants’ attitudes toward taking air taxis. 

\begin{table}[H]
\centering
\caption{Results of a quantitative analysis of the pre-and post-differences in ATJ’s attitudes to taking an air taxi.}
\label{tab:freq-7}
\begin{tabular}
{p{3.7cm}<{\centering} p{0.8cm}<{\centering} p{0.8cm}<{\centering} p{0.8cm}<{\centering} p{1.2cm}<{\centering} p{1.2cm}<{\centering}}\\ \hline
&N &Mean&SD & $t$ & $p$\\ 
\hline
Attitude (Before) &72 & $3.46$& $1.19$&& \\ 
 && && $-3.326$& $0.001$** \\ 

Attitude (After) & 72& $3.89$& $0.78$ & & \\ 
\hline
\end{tabular}
\begin{tablenotes}
    \footnotesize
    \item *$p\textless{}.05$; **$p\textless{}.01$; ***$p\textless{}.001$
    \end{tablenotes}
\end{table}

\begin{figure}[H]
  \centering
  \includegraphics[width=\linewidth]{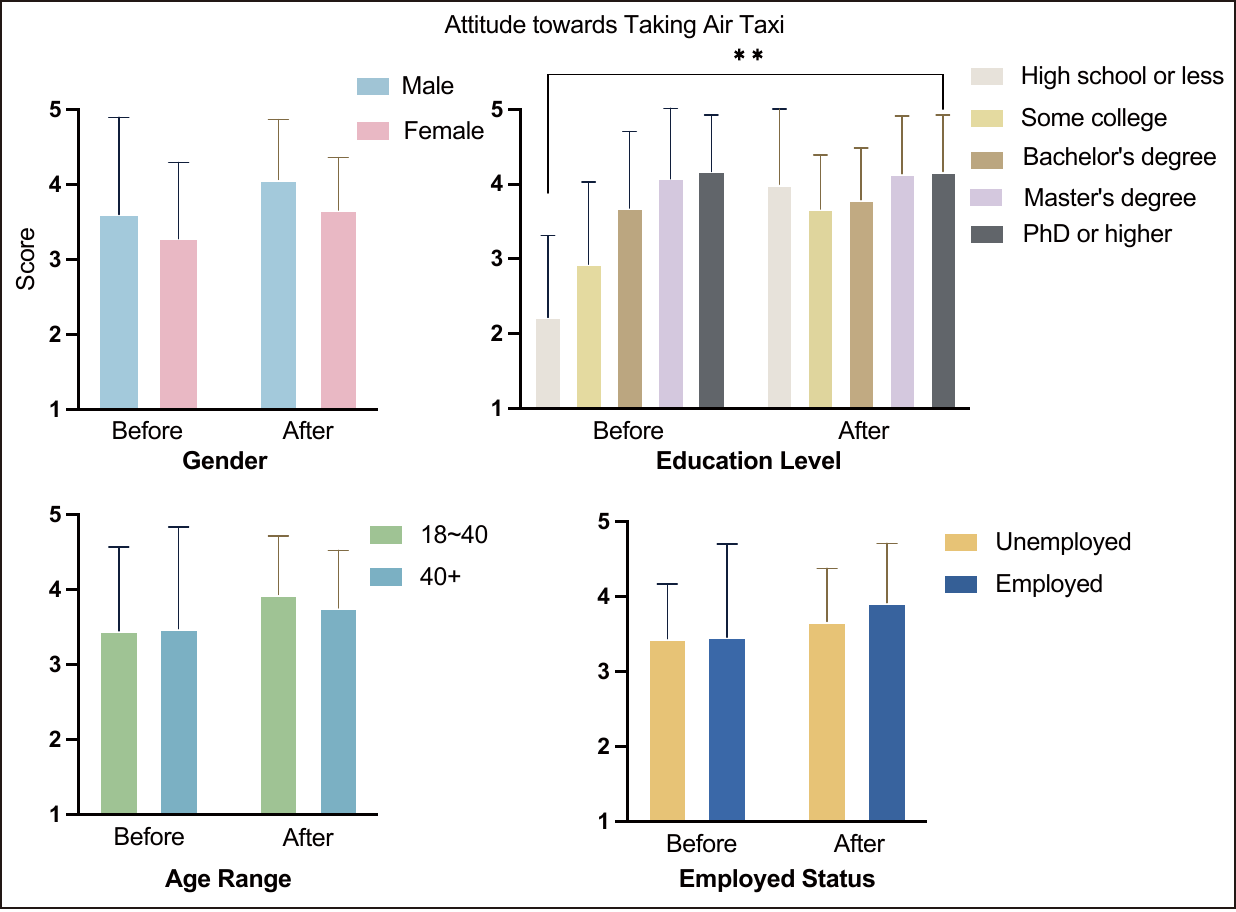}
  \caption{Results of a quantitative analysis of influencing factors affecting participants’ attitude change with ATJ. The four subfigures therein represent the different subgroups of participants in terms of gender, education level, age, and employed status. The vertical axis represents the mean of participants’ willingness to take an air taxi. *$p\textless.05$; **$p\textless.01$; ***$p\textless.001$}
  \label{fig:30}
\end{figure}

Subsequently, the factors influencing participants’ attitudes toward taking an air taxi from user properties, including gender, education level, age, and employment status, were analyzed. As shown in Figure~\ref{fig:30}, the change in participants’ attitude toward taking air taxi decreases as their education level rises, with the mean value of participants with a high school education increasing from $2.22$ to $4.00$ after viewing ATJ, while the mean value of participants with a Ph.D.~education toward taking air taxi is $4.17$ for them. A repeated test ANOVA showed that there was a statistically significant difference in the change in attitude towards taking an air taxi after viewing the ATJ among different education levels ($F=3.733$, $p=0.008$*). 

However, the results showed that there were no statistically significant differences in attitude changes towards taking an air taxi according to gender ($F=3.550$, $p=0.064$), age group ($F=0.124$, $p=0.726$), or employment status ($F=0.201$, $p=0.655$) after viewing ATJ. After viewing the ATJ, the willingness to take an air taxi increased from the mean value of $3.60$ to $4.07$ for males and from the mean value of $3.28$ to $3.66$ for females. The willingness to take an air taxi increased from the mean value of $3.45$ to $3.94$ for the 18$\sim$40 age group and from the mean value of $3.48$ to $3.76$ for the 40+ age group. The willingness to take an air taxi increased from the mean value of $3.44$ to $3.67$ for the unemployed status group and from the mean value of $3.46$ to $3.92$ for the employed status group. Although the results of A repeated test ANOVA showed that there was no significant difference in the attitude change of these three user properties before and after viewing the ATJ. However, the attitude of willingness to take an air taxi improved for all three user properties after viewing the ATJ. And both before and after ATJ viewing, the willingness to take an air taxi improved more positively for males than for females, and the employed status group becomes a little bit more positive than the unemployed status group. The specific results are shown in Appendix D Table~\ref{tab:freq-8}.

\subsection{The Satisfaction with ATJ}

To further validate the quality of the ATJ designed using the generative AI. We quantitatively analyzed the satisfaction of 72 participants with ATJ. After watching the ATJ, the participants were more satisfied ($\mbox{mean}=3.78$, $\mbox{SD}=0.77$). 

Subsequently, participants of different genders, education levels, ages, and employment statuses were categorized separately to quantitatively analyze the impact of different group characteristics on satisfaction with ATJ, as shown in Figure~\ref{fig:4}). Please see Appendix D and Table~\ref{tab:freq-20} for the specific data. 

For gender impact, the data showed that males ($\mbox{mean}=3.98$, $\mbox{SD}=0.83$) were more positively satisfied with the ATJ than females ($\mbox{mean}=3.53$, $\mbox{SD}=0.62$). An independent samples t-test showed a statistically significant difference in satisfaction with the ATJ according to gender ($t=2.508$, $p=0.014$*). 

Regarding the impact of educational level, the data revealed that participants’ satisfaction with the ATJ increased as their education level progressed, from a mean of $3.78$ for the high school degree to a mean of $4.17$ for the Ph.D.~degree. However, the repeated-test ANOVA results showed that there was no statistically significant difference in the satisfaction levels of participants with different education levels toward ATJ ($F=1.319$, $p=0.272$). 

Regarding the effect of the age group, the results showed that the 40+ group ($\mbox{mean}=3.98$, $\mbox{SD}=0.83$) scored slightly higher than the 18$\sim$40 group ($\mbox{mean}=3.81$, $\mbox{SD}=0.81$). However, Independent Samples t-test showed there was no statistically significant difference between the 18$\sim$40 and 40+ age groups in terms of satisfaction with Airbus Journey ($t=-0.222$, $p=0.825$). 

For the employment status impact, the results show that the unemployed group ($\mbox{mean}=3.78$, $\mbox{SD}=0.67$) was the same as the employed status group ($\mbox{mean}=3.78$, $\mbox{SD}=0.79$). An Independent samples t-test showed that there was no statistical difference between the unemployed and employed status on the satisfaction level with ATJ ($t=0.000$, $p=1$). 

\begin{figure}[H]
  \centering
  \includegraphics[width=\linewidth]{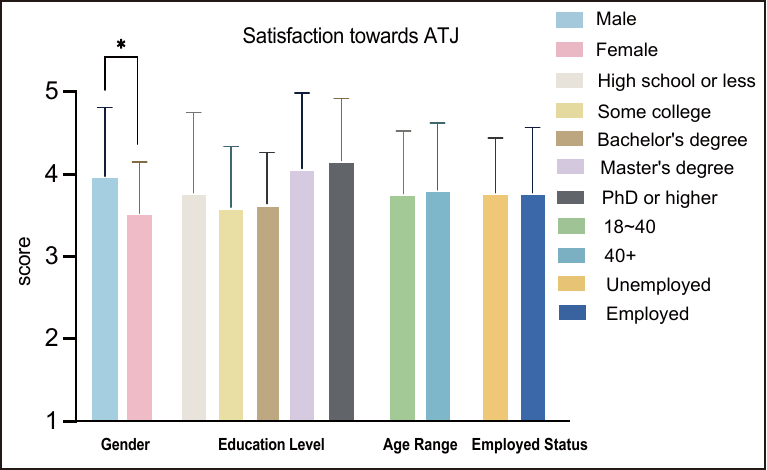}
  \caption{Results of a quantitative analysis of influencing factors affecting participants’ satisfaction with ATJ. The vertical axis represents the participants’ satisfaction ratings of ATJ. The horizontal axis shows the different subgroups of the participants in terms of gender, education level, age, and employment status. *$p\textless.05$; **$p\textless.01$; ***$p\textless.001$}
  \label{fig:4}
\end{figure}

\section{Discussion}

\subsection{User Study with LLM and AI Image and Video Generators}

In this study, we used an LLM and AI image and video generators to conduct an air taxi user study following the logic of design thinking. As the iterative steps of design thinking, the corresponding GPT-4 prompts were also designed, including “Prompt Empathize”, “Prompt Define”, “Prompt Ideate”, and “Prompt Prototype” (see Appendix B). After defining “Air taxi”, “Passenger”, and “Motivation” as ALOs, the critical components including the pain point, motivation, user need, design requirement, verbal mockup (see Appendix C - Table~\ref{tab:freq-5}), and storyboard (see Appendix C - Table~\ref{tab:freq-6}) were analyzed and generated progressively from GPT-4. The verbal mockup was designed with four modules: an app-based platform, real-time tracking, limited seating capacity air taxi, and AI customer support. The verbal storyboard of the simulated flight was generated with ten scenes to describe how the user interacted with the mockup. Therefore, GPT-4 can bring the researcher and designer new designs to the imagination of stakeholders in air taxi studies. In addition, the simulated flight generated from GPT-4 proved to be beneficial to the participants in improving their attitude toward taking air taxis and was satisfactory.

In this study, we used an LLM and AI image and video generators to conduct an air taxi user study following the logic of design thinking. As the iterative steps of design thinking, the corresponding GPT-4 prompts were also designed, including “Prompt Empathize”, “Prompt Define”, “Prompt Ideate”, and “Prompt Prototype” (see Appendix B). After defining “Air taxi”, “Passenger”, and “Motivation” as ALOs, the critical components including the pain point, motivation, user need, design requirement, verbal mockup (see Appendix C - Table~\ref{tab:freq-5}), and storyboard (see Appendix C - Table~\ref{tab:freq-6}) were analyzed and generated progressively from GPT-4. The verbal mockup was designed with four modules: an app-based platform, real-time tracking, limited seating capacity air taxi, and AI customer support. The verbal storyboard of the simulated flight was generated with ten scenes to describe how the user interacted with the mockup. Therefore, GPT-4 can bring the researcher and designer new designs to the imagination of stakeholders in air taxi studies. 

To prove the validity of the air-taxi design idea generated by GPT-4, we conducted a user test to evaluate the mockup and the storyboard. Therefore, we recruited 72 people for the user test and designed a questionnaire to evaluate the simulated flight. The simulated flight from Scenes 1 to 10 was visualized using Midjourney and Runway (in Figure~\ref{fig:8}). The mockup of the air taxi UIs was visualized by a Midjourney. Storyboard scenes 8$\sim$10 were generated using Runway. Finally, the virtual experimental scenarios for the air taxi study was built using GPT-4, Midjourney, and Runway. The participants were then asked to view the verbal and visual simulated flights to evaluate their interactions with the air taxi. The results of the user test, as presented in Table~\ref{tab:freq-7} in Section $4.1$, show a significant positive difference in the participants’ attitudes towards the air taxi after viewing the simulated flight generated by the generative AI ($p=0.001$*). The participants rated their satisfaction with the simulated flight highly ($\mbox{mean}=3.78$, $\mbox{SD}=0.77$). These results demonstrate that our method can be used to build air taxi UIs and UX that satisfy users. This can assist air taxi developers in conducting user studies during the early design phase.

We demonstrated the potential of using generative AI in virtual environmental scenarios to conduct a emerging and unfamiliar user study such as air taxi, which face limitations, including safety concerns and iterative efficiency. The related studies typically recruited a large number of participants to facilitate comprehensive research on air taxis \parencite{al2020factors, Surveydata, ward2021safety,rice2022vertiport}. In contrast, through our designed GPT-4 prompts, these studies could reduce the effort required to conduct empathy user interviews, making our user study less demanding in terms of the number of participants. Furthermore, conducting these user studies using the prompts, a field yet unexplored by public conceptual research and the absence of developer experience could provide stakeholders with a suitable design solution in the early stage of the design to help them conceptualize the object. In contrast to researchers who dedicate substantial effort to creating real or virtual environmental scenarios for user experiments, our approach leverages AI image and video generators to produce necessary materials for user tests more efficiently and cost-effectively. Rather than participants going through a real user experiment, having participants experience a user-tested virtual experimental environment generated by our prompts and AI image and video generators also avoids the high-risk nature in the experiments. As such, our approach simplifies the establishment of the experimental scenarios for these study in the user study process, making it more cost-effective, efficient, and safe. This provides creativity and imagination for an user study that is high-tech and unprecedented in early design phase. 

However, a few limitations regarding the interaction with participants. To address the limitations of participant engagement, which relied solely on video depictions of the interactions between participants and air taxis, we propose an enrichment of the user-testing methodology. Hence, we aim to enhance user testing by introducing more physical interactions with the virtual experimental scenarios for the participants.  We intended to advance from the current offering of only textual descriptions and videos in the questionnaire section to provide participants with a more dynamic and immersive experience. 

Further study may explore studies beyond air taxis, and it might be useful in space exploration area and autonomous driving. These could be concentrated in high-tech areas, such as service robots, which signify high investment, high risk, and lengthy exploration periods, yet remain largely unfamiliar and less accepted by the average user. To enhance the comprehensiveness of our work, we aim to include two additional case studies in other fields, following our established methodology. We anticipate that our generative AI-driven approach may affect and contribute to user study in various domains. 

\subsection{Customer Experience Design for Air Taxi}

According to the user tests, the Air Taxi Journey (ATJ) designed by Generative AI performed satisfactorily. The mockup of the air taxi was proposed with four modules: an app-based platform, real-time tracking, limited seating capacity, and AI customer support. For real-time tracking, we propose designing a GPS and status update alert service. For the limited seating capacity air taxi, we suggest implementing a reasonable seating arrangement, safety equipment such as seat belts, and adherence to relevant regulations. Finally, for AI customer support, we recommend designing a typing box and voice service for communication with customer service, as well as a service to assist passengers with bookings and cancellations. Additionally, a suggestion box should be included. It is also desirable for the service to be available on multiple communication terminals.

The impact of different factors on the willingness and satisfaction of people with air taxis was also analyzed. Regarding the willingness to take an air taxi, our experimental results show that the ATJ generated from the generative AI strengthened the willingness to take an air taxi. Our discovery has contributed to the development of air taxis because some studies indicate public hesitancy and low trust levels toward air taxis and autonomous aircraft \parencite{hughes2009automated,mehta2017perceptions, rice2014passengers, rice2015depression}. Nevertheless, the acceptance of users and society is crucial in developing new technologies \parencite{otte2018user, yavas2023consumer}. The acceptance of air taxis as a new and unpublished technology will influence its development. Therefore, it might be possible in the future to use the results of user studies conducted by the generative AI to enhance users’ positive attitudes towards new technologies such as air taxis. 

Furthermore, there was a statistically significant difference ($p=0.008$*) in the willingness to take an air taxi between the different levels of education (see Table~\ref{tab:freq-8}). The results showed that simulated flights had the greatest effect on the group with low educational qualifications and had almost no effect on the group with the highest educational qualifications. This study also found significant differences in the acceptance of air taxis for knowledge \parencite{ward2021safety}. Moreover, equally relative to emerging transportation studies, individuals with higher education favor autonomous vehicles \parencite{haboucha2017user}. In addition, studies have explored how familiarity with autonomous flight, fun, wariness of new technology, happiness, fear, age, and education level influence consumers’ willingness to fly in autonomous vehicles \parencite{rice2019factors, ward2021safety}. Education is also mentioned as a factor influencing people’s willingness to use air taxis. Therefore, when considering the promotion of air taxis, the market should focus on the education factor in designing promotions or customer segmentation. For instance, to make transportation modes such as air taxis accessible and bring benefits to more people, the promotion of air taxis can focus on people with low education levels. 

Regarding the different factors affecting the satisfaction of people with simulated flights, it was found that there was a statistical difference in the level of satisfaction with the ATJ by gender (see Figure~\ref{fig:4}). The study indicates that demographics, such as gender, influence the UX toward autonomous cars \parencite{rodel2014towards}, as well as a newer mode of transport. In a survey on satisfaction with public transport, it was found that satisfaction with public transport had very different impacts on male and female groups, and it is necessary to increase efforts to improve females' satisfaction with public transport \parencite{fu2017exploring}. Females have a distinct need for safety and security owing to their fear of certain transit environments \parencite{loukaitou2014fear}. Consequently, gender differences should be considered in air taxi UX design and study to allow more populations to enjoy air taxis. Our study and its results provide a good reference for air taxi service developers when considering user preferences.

\subsection{The Possibility of LLM Simulating Participant Response}

To explore the extent of the LLM’s ability to conducting user studies with virtual user and virtual experimental scenario, 72 GPT-4 and Gemini participants’ responses were simulated, as shown in Figure~\ref{fig:1-1} (a) [IV]. This was to investigate the possibility that the LLM can simulate evaluation responses from participants. 

\subsubsection{Procedure: Real Responses from LLM Participants}

In Section $3.2.2$, 72 participants were recruited to evaluate the simulated journey. Here, we simulated the 72 participants' responses in GPT-4 and Gemini to the same questionnaire from real participants. 

Regarding the GPT-4 participants, the content of the questions and options in the questionnaire and key images of 10 scenes (see Figure~\ref{fig:8}) were provided to GPT-4. For Gemini participants, the content of the questions and options in the questionnaire and videos of the 10 scenes were provided. 

Next, 72 questions with personal information about 72 characters were asked in GPT-4 and Gemini, respectively. The question was, “\textit{If you are participant 1, who is age (20-30), nationality (Asian or Pacific Islander), gender (Male), education (Bachelor’s degree), employment status (Employed - Professional / White - Collar ($e.g.$, corporate, managerial, administrative)), personal interests (entertainment ($e.g.$, Traveling, Gaming)) and preference for transport (Private Car)}”. Finally, responses from GPT-4 and Gemini were generated.

\subsubsection{Effect of ATJ on the User}

We quantitatively analyzed the attitudes of 72 GPT-4 and Gemini participants toward air taxis after viewing the ATJ. As shown in Table~\ref{tab:freq-22}, after viewing the ATJ, the mean value of the 72 GPT-4 participants’ attitudes towards taking air taxis increased from $3.74$ to $3.87$. A paired-sample t-test revealed a statistically significant difference in the attitudes towards air taxis among the GPT-4 participants after viewing ATJ ($t=-3.384$, $p=0.001$*). Regarding the Gemini participants’ responses, their attitudes towards taking air taxis also increased after viewing ATJ, from $4.07$ to $4.44$. A paired samples t-test indicated that there was a statistically significant difference in Gemini participants’ attitudes towards taking air taxis after viewing ATJ ($t=-5.376$, $p\textless{}0.001$*) as well. 

For both GPT-4 and Gemini participants, there was a significant positive difference in the GPT-4 ($p=0.001$) and Gemini ($p\textless{}0.001$) simulated participants’ attitudes toward the air taxi after viewing the simulated flight generated by the generative AI. This is consistent with the responses from real user tests where we evaluated the impact of a design on users. This means that air taxi developers can collect basic information about their target users and simulate user evaluation responses from these target users regarding the effects of this design through LLM. 

\begin{table}[H]
\caption{Results of a quantitative analysis of the pre- and post-differences in ATJ’s attitudes towards taking the air taxi. This includes virtual response from the GPT-4 and Gemini.}
\label{tab:freq-22}
\resizebox{\textwidth}{!}{
\begin{tabular}{cccccccc}
\hline
&N & \multicolumn{2}{c}{Mean}& \multicolumn{2}{c}{SD} & $t$ & $p$\\ 
\hline
\multirow{2}{*}{}& \multirow{2}{*}{} & \multicolumn{1}{m{2cm}}{\centering Attitude (Before)} & \multicolumn{1}{m{2cm}}{\centering Attitude (After)} & \multicolumn{1}{m{2cm}}{\centering Attitude (Before)} & \multicolumn{1}{m{2cm}}{\centering Attitude (After)} & & \\
\hline
\begin{tabular}[c]{@{}c@{}}GPT-4 \end{tabular} & 72& $3.74$& $3.87$& $0.99$& $1.05$ & $-3.384$& $0.001$** \\ 

\begin{tabular}[c]{@{}c@{}}Gemini \end{tabular} & 72& $4.07$& $4.44$& $0.35$& $0.50$&$-5.376$&$\textless{}0.001$*** \\ 
\hline
\end{tabular}
}
\begin{tablenotes}
    \footnotesize
    \item GPT-4: GPT-4 Participants
    \item Gemini: Gemini Participants
    \item *$p\textless{}.05$; **$p\textless{}.01$; ***$p\textless{}.001$
    \end{tablenotes}
\end{table}

\subsubsection{Differences in the Each Questions}
\begin{figure}[H]
  \centering
  \includegraphics[width=\linewidth]{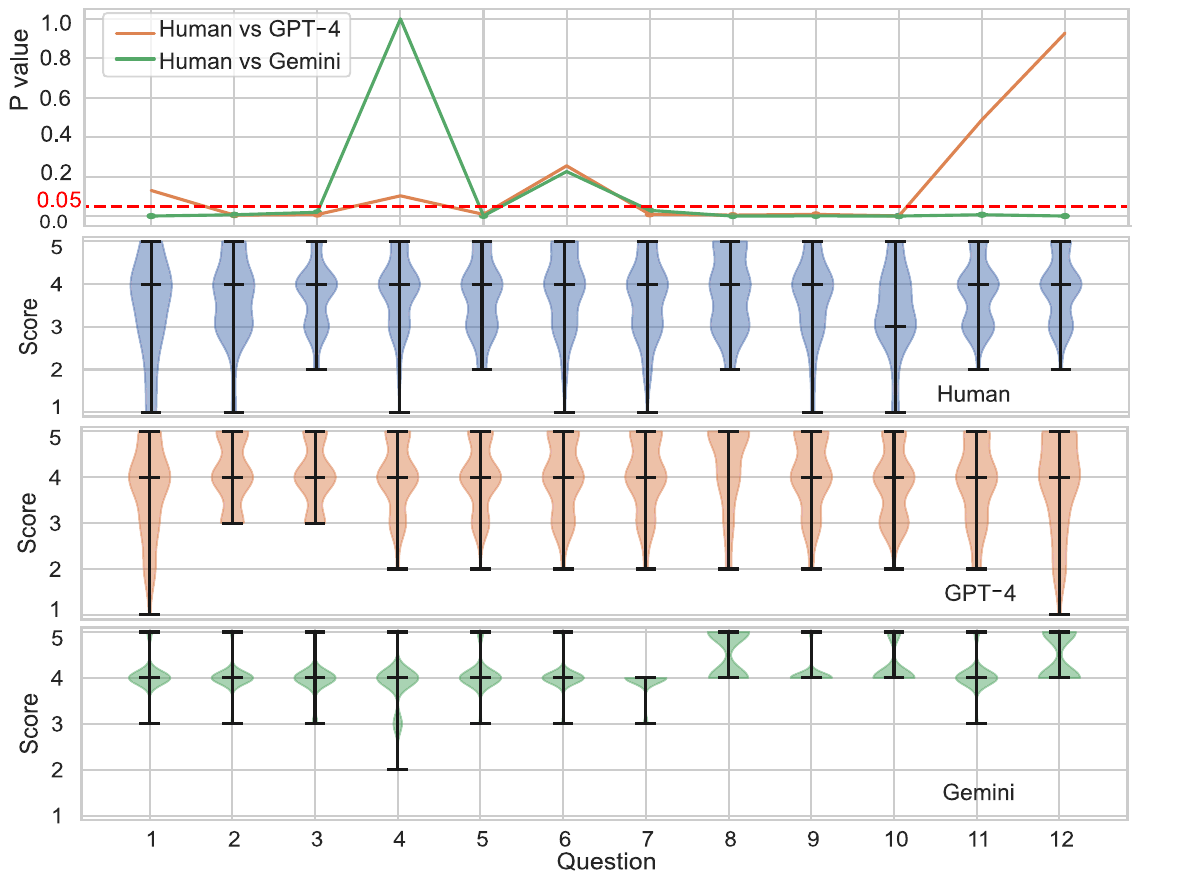}
  \caption{Mean and p-value of 72 real and 72 virtual responses for 12 questions from Q1 to Q12. The vertical axis represents the mean score and p-value, and the horizontal axis represents the 12 questions, *$p\textless{}.05$; **$p\textless{}.01$; ***$p\textless{}.001$}
  \label{fig:9}
\end{figure}

However, when we compared the differences between the 72 real and virtual participants as a population on the 12 questions, there were some differences between the responses of the real and virtual participants (see Figure~\ref{fig:9}). 

Regarding the GPT-4 emulated participants’ responses, paired-sample t-tests showed that only Q1, Q6, Q11, and Q12 had no significant differences in the ratings of real and GPT-4 emulated participants’ responses to the 12 questions, while the remaining eight questions had significant differences. Regarding the Gemini emulated participants’ responses, the paired-samples t-test showed that only Q4 and Q6 had no significant differences in the ratings of real and Gemini emulated participants’ responses to the 12 questions, whereas the remaining 10 questions had significant differences. Although both models represent advances in the LLM for simulating human responses to transportation, GPT-4 outperforms Gemini. This is consistent with the results in the comparison of LLM in education and healthcare applications, that is, GPT-4 has higher performance and diagnostic accuracy than Gemini for multimodal applications respectively \parencite{lee2023gemini,pal2024gemini,carla2024exploring}. 

However, there were no significant differences between the real and virtual participants’ responses to some of the questions. Further studies are essential for understanding how a group of virtual participants can correctly predict the outcomes of real participants. This effect can potentially be explained by the “wisdom of the crowd” or “diversity prediction theorem” as postulated by Scott E. Page \parencite{page2008difference}. Another study found that it is possible to reduce the reliance on human-generated data after self-training in response to the LLM \parencite{singh2023beyond} as well. The diversity of the virtual participants may have contributed to the prediction of correct answers. While this is still at the hypothesis stage, the results indicate the possibility of using LLM for a user study, given that enough virtual participants are used in the process. Therefore, our study shows that it is possible to predict participants’ responses by simulating real participants in the user test of an air taxi using the GPT-4 before the real user test. The results of the simulation can be used as a reference for further refinement before conducting real user tests.

The limitation is that the performance of virtual humans did not fully align with that of real human responses in certain questions. Further studies are essential for understanding how a group of virtual participants can correctly predict the outcomes of real participants. Consequently, future research should be dedicated to delving deeper into methodologies aimed at enhancing the capacity of LLM to emulate responses similar to those observed in real-world scenarios. We believe that it will be possible to use the LLM to simulate more convincing participant responses to conduct user tests in the future. This would make user experiments more inclusive and safer, as it allows researchers to better run user studies with wider groups of people and also reduces the risks to the participants.

\section{CONCLUSION}

This study explored the ability of generative AI to conduct user studies in emerging and unfamiliar contexts, using air taxis as a case study. Real user evaluations confirmed the efficacy of the proposed approach, demonstrating an increased willingness to use air taxis. The analysis revealed that educational level significantly influenced user attitudes, while gender affected satisfaction, with females being more selective and less satisfied with the air taxi experience than males. These findings suggest that future air taxi designs and marketing should consider educational backgrounds and gender. We also explored that the the potential of LLM to emulate the participant response and capturing the individual information of the targeted user. Although there are still some limitations in narrowing the gap between real and virtual responses, our current study proves that the LLM can simulate some of the responses to user evaluations and thus predict the outcome. 

In conclusion, our generative AI-based approach to creating virtual user experiences offers a feasible and insightful method for improving user studies, particularly in contexts with safety and iterative efficiency constraints, such as air taxis. The virtual environmental scenarios mitigate safety concerns of real environments in the early design stages. Additionally, it leverages the power of LLM and AI image and video generators to facilitate rapid iterative design processes, enhancing real vehicle design by closely approximating virtual evaluation results to real-world outcomes. In the future, we aim to improve the use of the generative AI approach and better simulate participant responses to extend our methodology to various user study areas. We believe our study represents a significant step towards more safe and effective user studies.

\section*{Disclosure statement}

The authors report there are no competing interests to declare.

\section*{Funding}

This work was supported by the Support for Pioneering Research Initiated by the Next Generation (SPRING) under Grant number JPMJSP2124.


\printbibliography


\appendix

\section{Questionnaire}
\subsection{Consent Form}

Air Taxi Journey

I received a full explanation of this study, “Evaluation of user experience simulation in a virtual environment using LLM (GPT-4) toward air taxi”, including an overview of the study, the need for experimental collaborators, methods, safety considerations and protection of personal data. During the explanation, it was also explained to me that I would not suffer any disadvantages if I did not agree to cooperate in this experiment and that even after giving my consent, I could withdraw at any time of my own free will without suffering any disadvantages. 
 
The aim of the project is to conduct user experiments in a virtual environment without real people, for example, to simulate an air taxi service. By automating user experiments without real participants, for example, dangerous experiments and burdensome tasks can be avoided. This shows that LLM can be used to carry out user research without actual people.

The questionnaire was voluntary and participants could refuse or withdraw from answering the questionnaire at any time. The questionnaire will take approximately 15~20 min. The results of the experiment may be used in documents and papers published inside and outside the University, but in all cases, the privacy of the participants will be protected. 

1. I understand the above and agree to cooperate with the questionnaire. \\
Yes\\
No

\subsection{Personal Information of Participants}

Hi! Nice to me you! Let's introduce each other! \\
 
To ensure that we gather accurate insights, could you kindly provide some information about yourself?\\ 
2. What is your age range? \\ 
Below 20,
20-30,
31-40,
41-50,
51-60,
Above 60

3. Which of the following best describes you?
Please select one answer. 

Asian or Pacific Islander, 
Black or African American, 
Hispanic or Latino, 
Native American or Alaskan Native, 
White or Caucasian, 
Multiracial or Biracial, 
Other (please specify)

4. What is your gender?
Male,
Female,
Non-binary,
Prefer not to say

5. What types of hobbies or leisure activities do you regularly participate in?
(multiple choice available) \\
- Physical Activities (e.g., Sports, Outdoor Activities), \\
- Creative Pursuits (e.g., Art, Music, Cooking), \\
- Intellectual Activities (e.g., Reading, Tech-related), \\
- Entertainment (e.g., Traveling, Gaming), \\
- Other (please specify)\\

6. What is your current employment status?
(multiple choice available) \\
Employed- Professional/White-Collar (e.g., corporate, managerial, administrative)\\
Employed- Service Industry (e.g., hospitality, retail, tourism)\\
Employed- Technical/Blue-Collar (e.g., trades, labour, manufacturing)\\
Employed- Educational (e.g., teaching, training)\\
Employed- Medical/ Healthcare\\
Employed- Self-Employed/Business Owner\\
Employed- Arts/Creative (e.g., design, music, writing)\\
Employed- Information Technology\\
Unemployed\\
Retired\\
Student\\
Other (please specify)\\

7. Which best describes your level of education? \\
- High school or less\\
- Some college\\
- Bachelor's degree\\
- Master's degree\\
- Ph.d.~ or higher\\

8. When you feel that your destination is relatively far away for you, what mode of transport do you usually choose?
(multiple choice available) \\
- Private Car\\
- Taxi Car\\
- Public transportation\\
- Bicycle\\
Other (please specify)\\

9. If an air taxi service is launched, how probable is it that you will take it? \\
Strongly impossible,	impossible,	neutral	possible,	Strongly possible

\subsection{Let's enjoy the air taxi journey}

Alright, now it's time to hop on our virtual air taxi journey!\\
This journey is split into 10 different scenes, each with its own unique experience.
After each scene, we'd love to hear your thoughts with just 1 or 2 quick questions.
Your input really helps us out!\\

10.  “Scene1. Passenger opens the app and logs in to their profile.”\\

\begin{figure}[H]
  \centering
  \includegraphics[width=\linewidth]{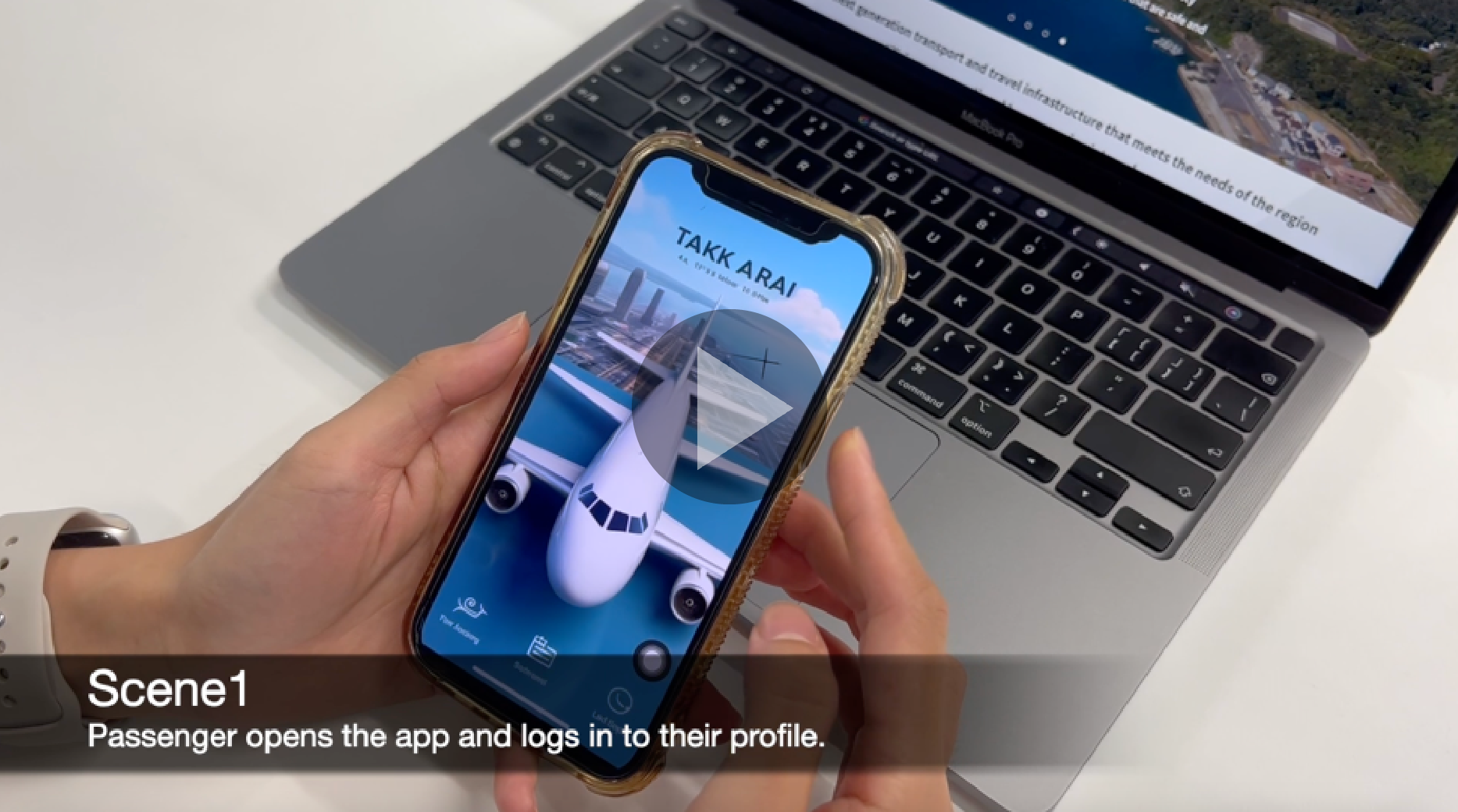}

  \label{fig:101}
\end{figure}
How do you like the initial interaction with the app from open the app to log in profile? \\
Strongly dislike,	dislike,	neutral,	like,	Strongly like\\
11. “Scene2. Passenger navigates to the booking section and inputs their travel requirements: from Tsukuba to Tokyo.”\\

\begin{figure}[H]
  \centering
  \includegraphics[width=\linewidth]{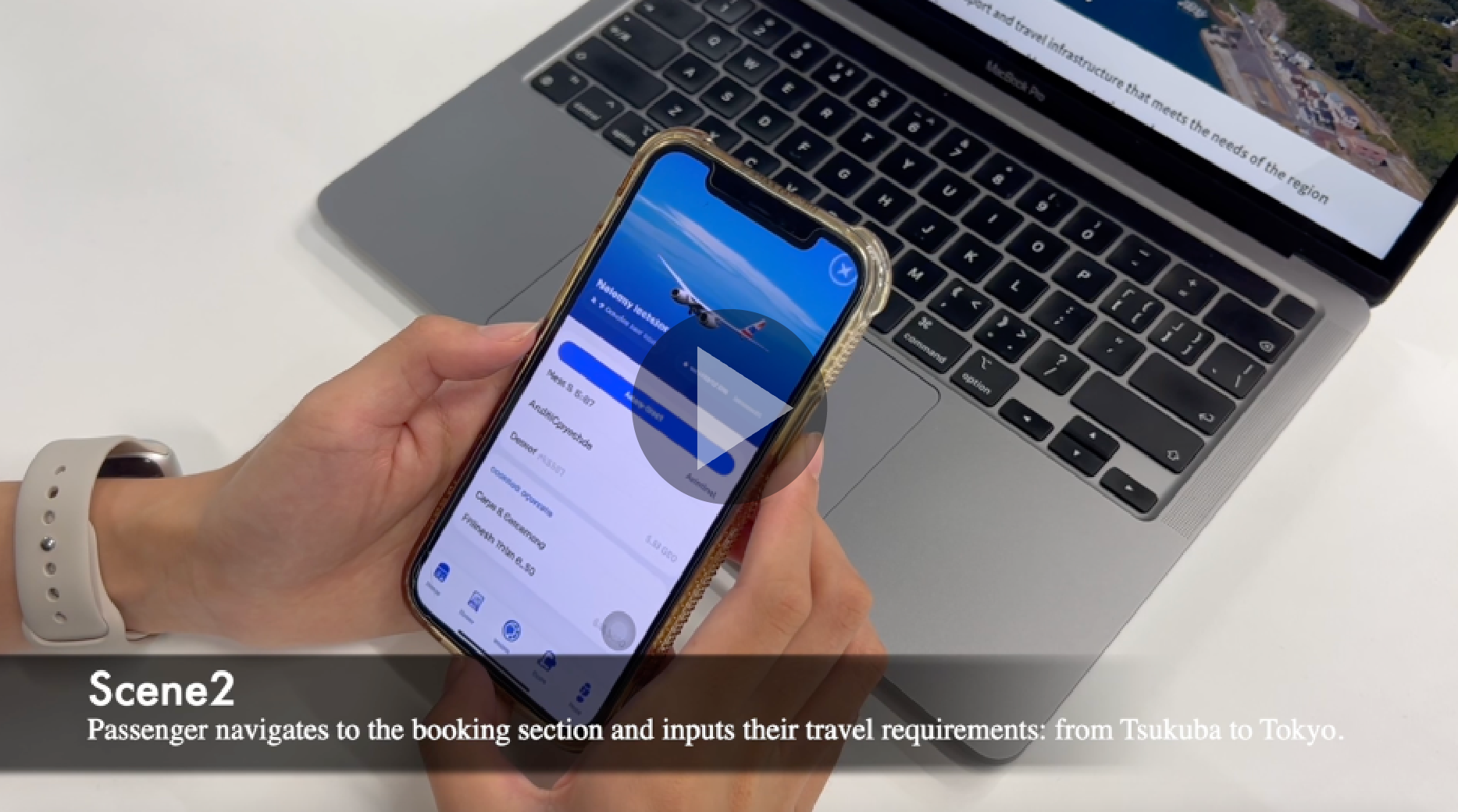}

  \label{fig:101}
\end{figure}
How do you feel about the navigation in the booking section and entering travel requests in the app? \\
Strongly dislike,	dislike,	neutral,	like,	Strongly like\\
12. "Scene3. The app shows a list of available air taxi with their respective times and prices."\\

\begin{figure}[H]
  \centering
  \includegraphics[width=\linewidth]{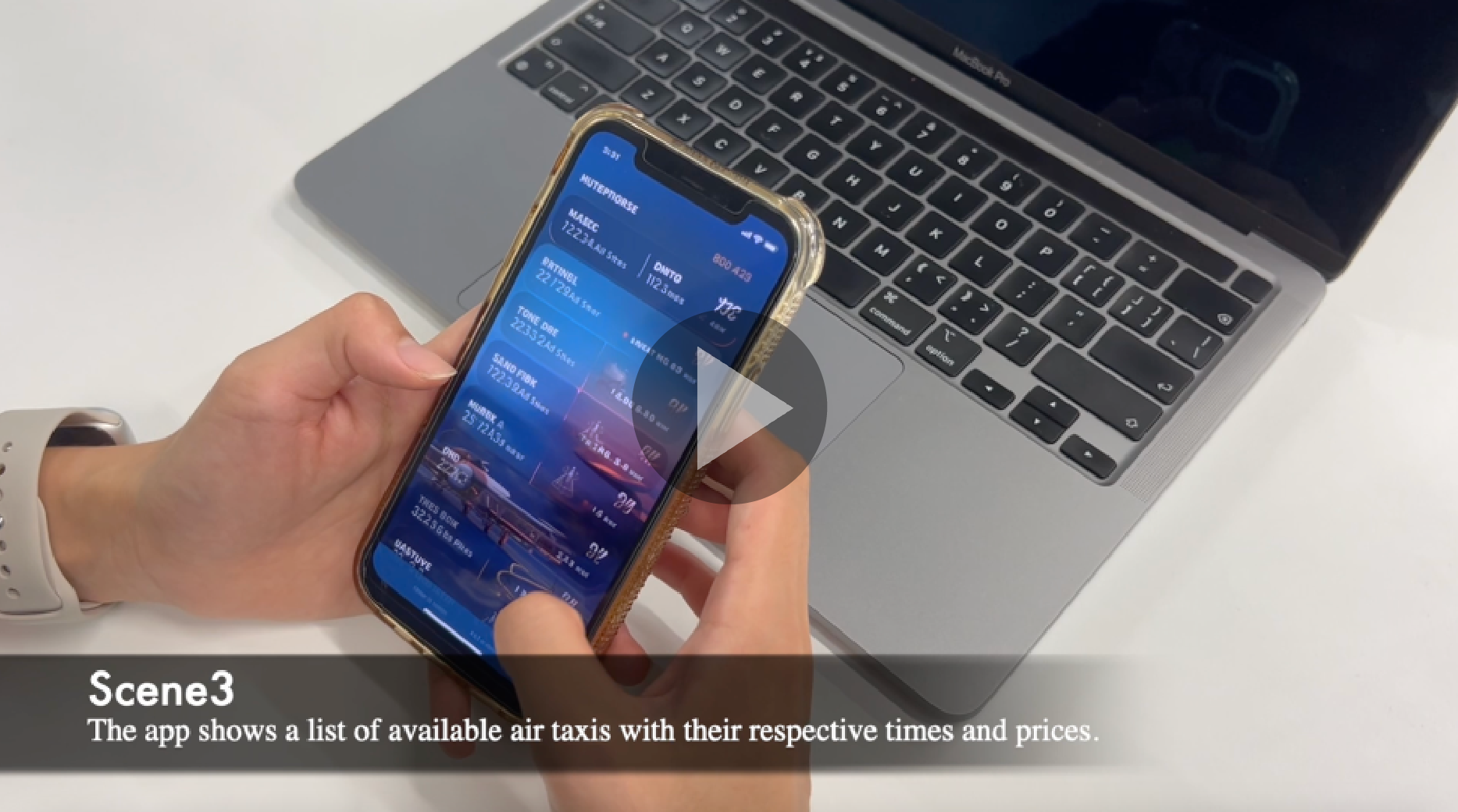}

  \label{fig:101}
\end{figure}
Do you feel the app provided sufficient information (times and prices) for you to make an informed decision? \\
Strongly insufficient,	insufficient,	neutral,	sufficient,	Strongly sufficient\\
13. "Scene4. Passenger selects their preferred air taxi and time and proceeds to the payment section."\\

\begin{figure}[H]
  \centering
  \includegraphics[width=\linewidth]{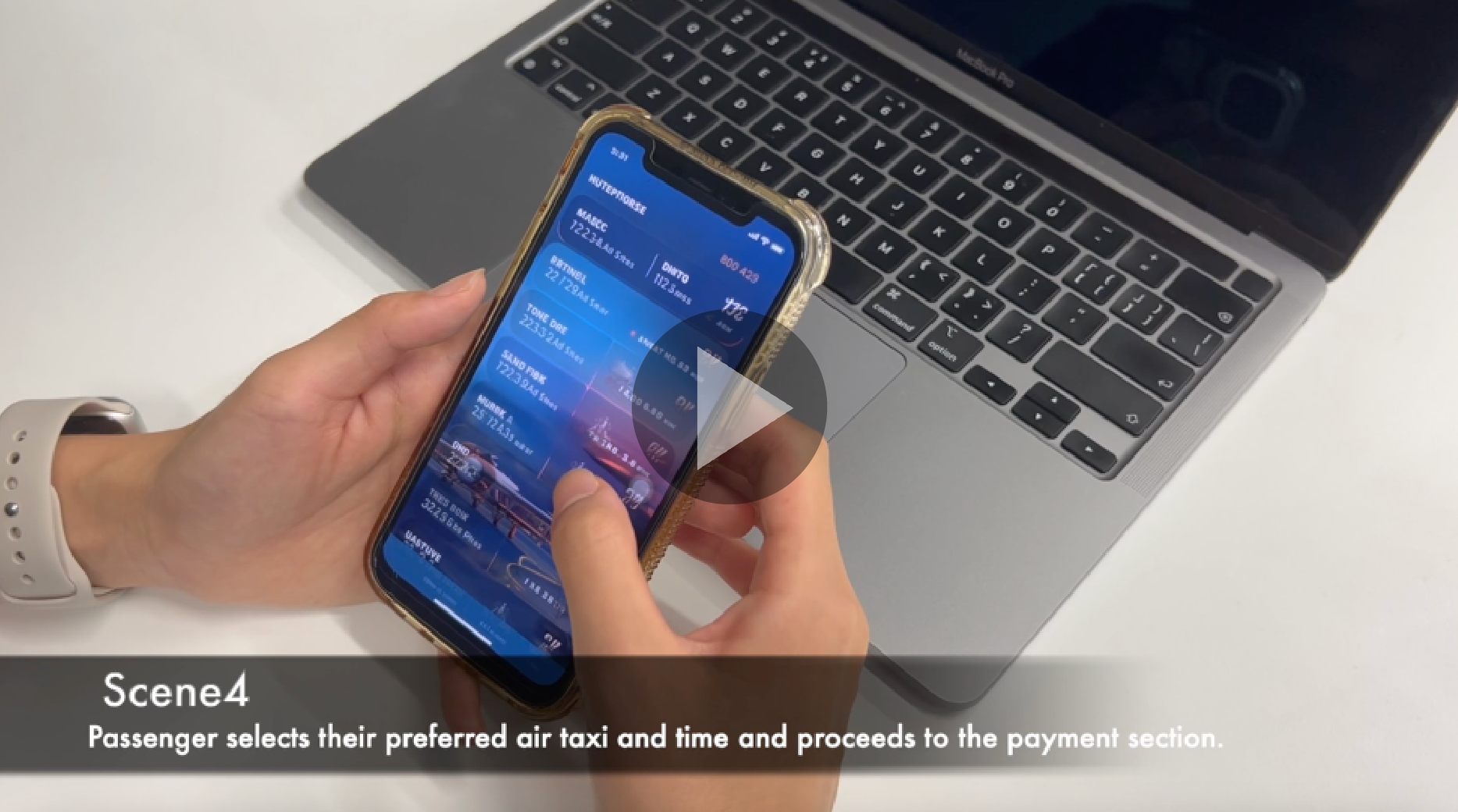}

  \label{fig:101}
\end{figure}
How easy is it for you to choose a preferred air taxi and time? \\
Strongly uneasy,	uneasy,	neutral,	easy,	Strongly easy\\
14. "Scene5: Passenger confirms their payment through a secure payment method, like PayPal."\\
\begin{figure}[H]
  \centering
  \includegraphics[width=\linewidth]{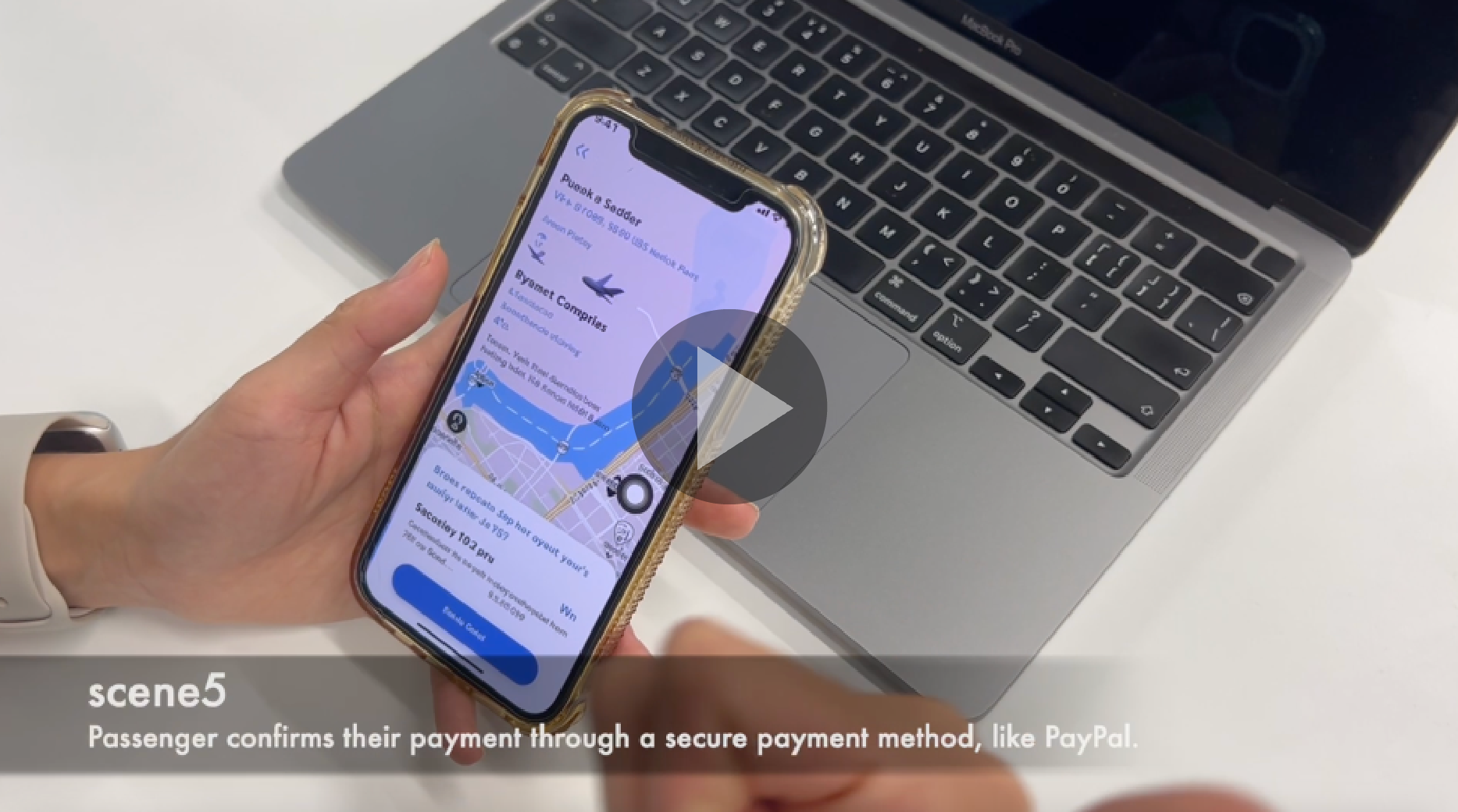}

  \label{fig:101}
\end{figure}
How esay does the payment process look? \\
strongly uneasy,	uneasy,	neutral,	easy,	strongly easy\\	
15. "Scene6. The booking is confirmed, and Passenger gets a confirmation message with details about the air taxi's departure time and location."\\
\begin{figure}[H]
  \centering
  \includegraphics[width=\linewidth]{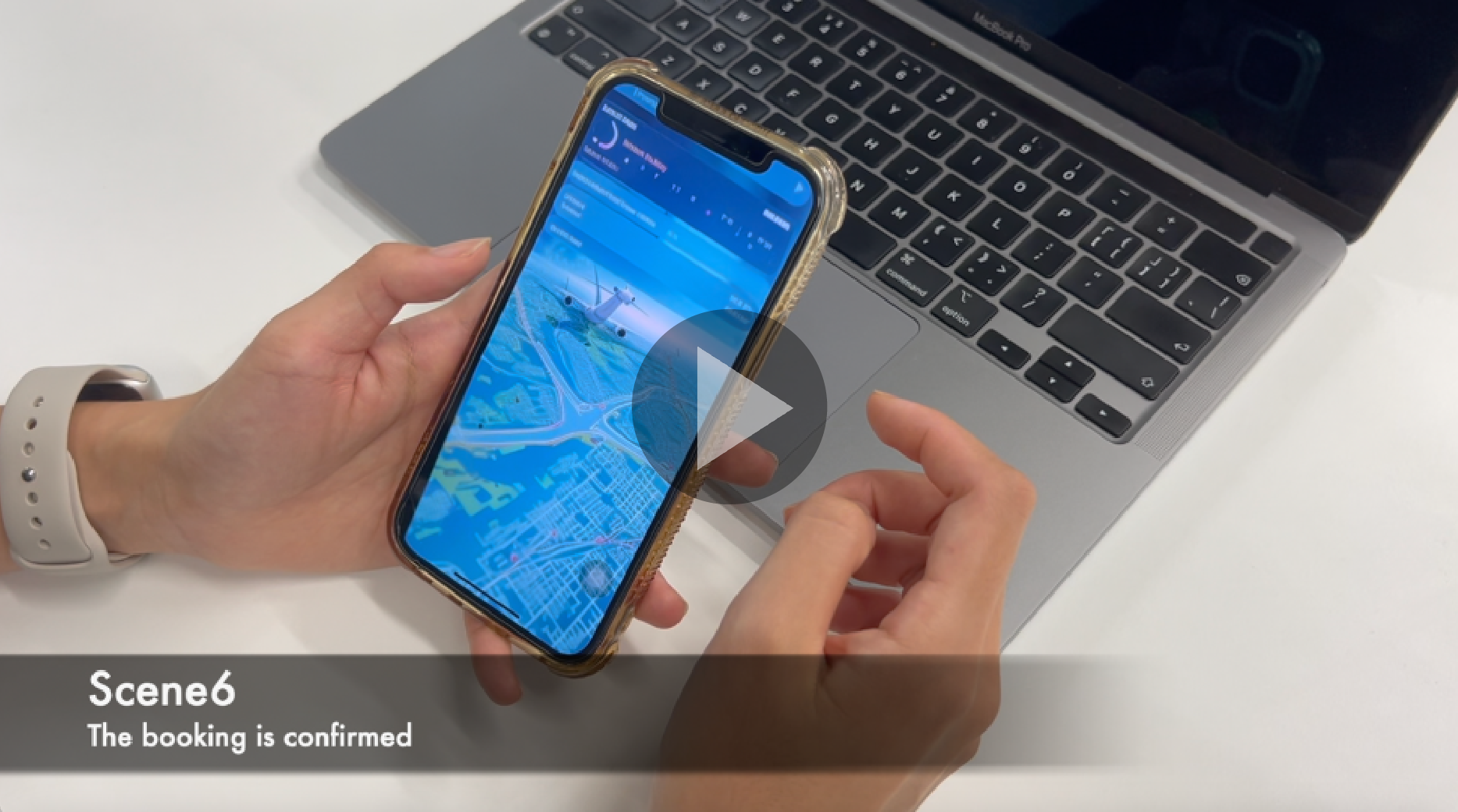}

  \label{fig:101}
\end{figure}
Do you feel the confirm message provided sufficient information (the air taxi's departure time and location) for you? \\
Strongly insufficient,	insufficient,	neutral,	sufficient,	Strongly sufficient\\
16. "Scene7. On the day of the departure, Passenger receives real-time updates on the air taxi's location and schedule through push notifications."\\
\begin{figure}[H]
  \centering
  \includegraphics[width=\linewidth]{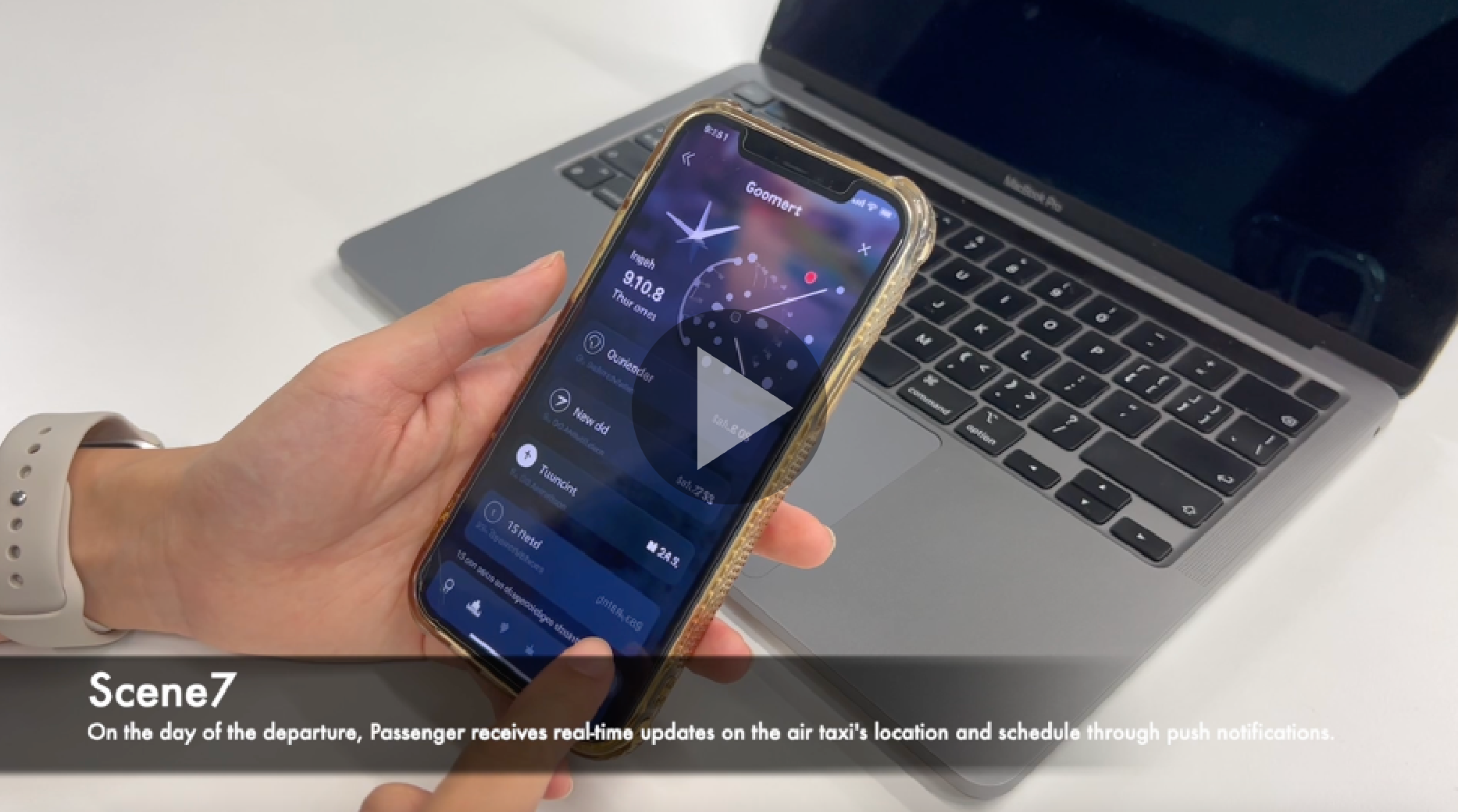}

  \label{fig:101}
\end{figure}
How beneficial would real-time updates on the air taxi's location and schedule be on your departure day? \\
Strongly useless,	useless,	neutral,	useful,	Strongly useful\\	
17. "Scene 8. Passenger arrives at the departure point and boards the air taxi. They find a comfortable seating arrangement and safety features, such as seat belts and emergency equipment." \\
\begin{figure}[H]
  \centering
  \includegraphics[width=\linewidth]{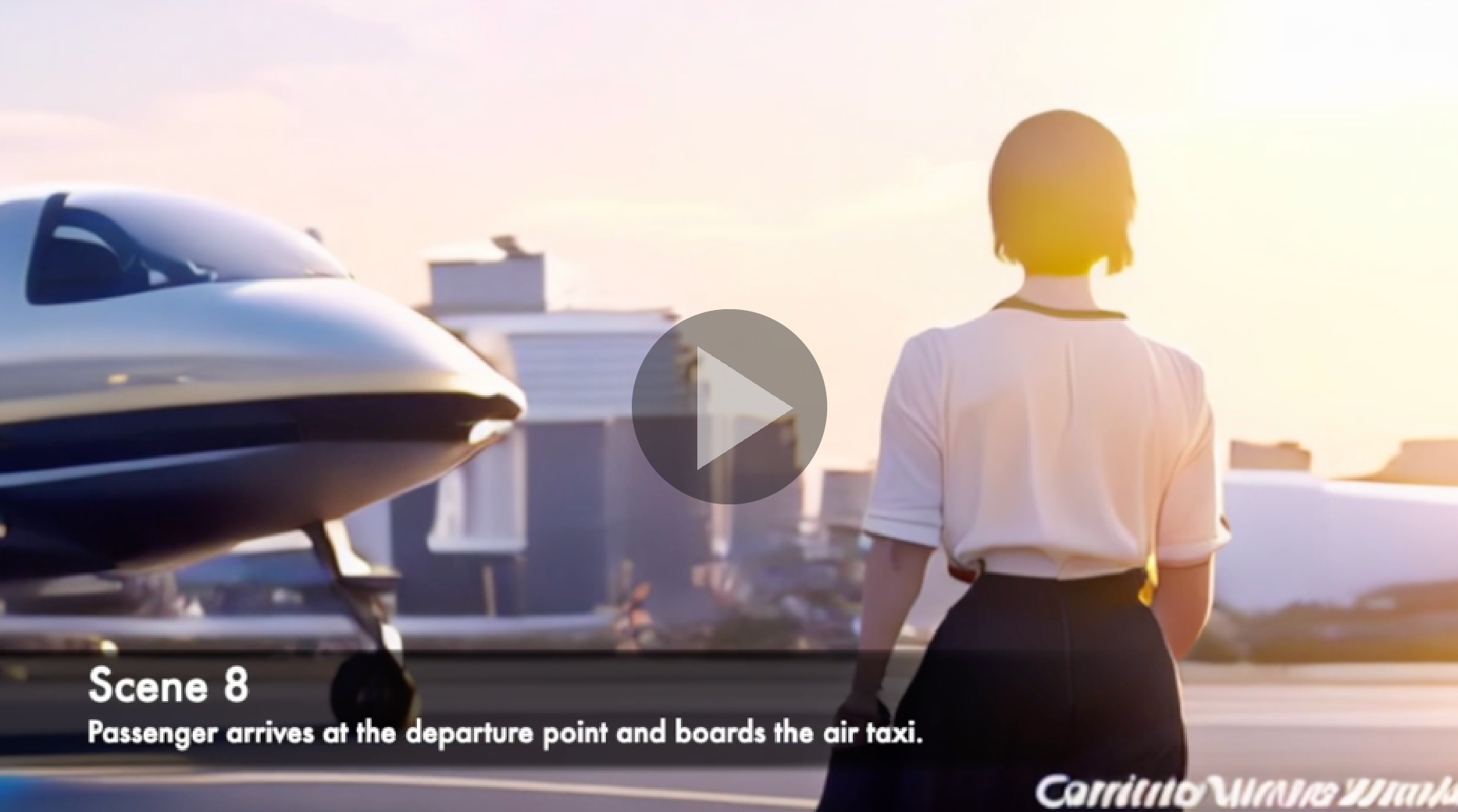}

  \label{fig:101}
\end{figure}
What are your perceptions of comfort inside an air taxi?\\
strongly Uncomfortable,	Uncomfortable,	neutral,	comfortable,	strongly comfortable
 \\
18. "Scene9. During the flight, Passenger has a question about the air taxi etiquette. They use the app to contact AI Customer Support, which promptly responds to their query." \\	
\begin{figure}[H]
  \centering
  \includegraphics[width=\linewidth]{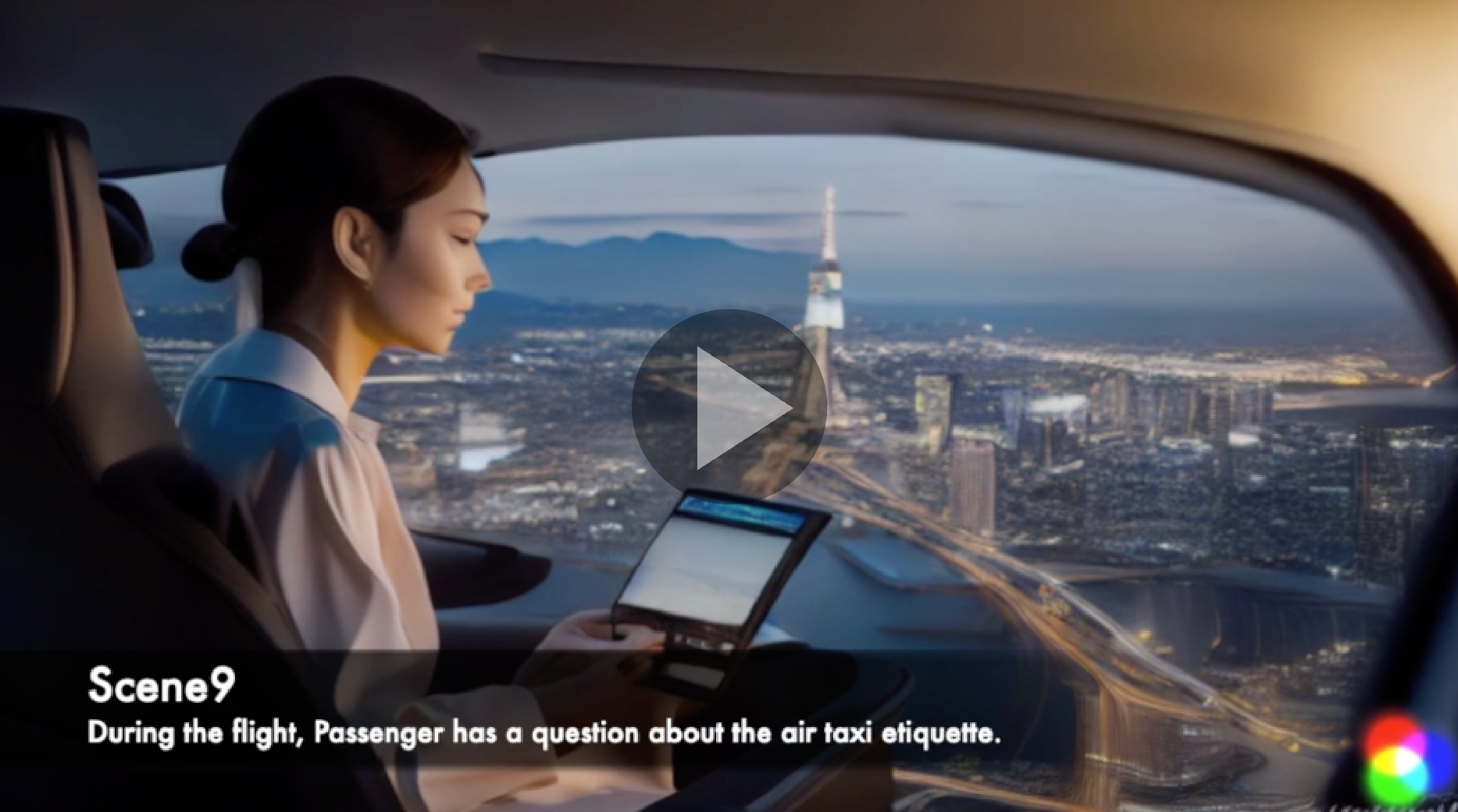}

  \label{fig:101}
\end{figure}
Do you feel that the AI Customer Support function has been helpful?\\
Strongly unhelpful,	unhelpful,	neutral,	helpful,	Strongly helpful\\
19. "Scene10. Passenger arrives in Tokyo, and the Air Taxi Journey is concluded."\\

\begin{figure}[H]
  \centering
  \includegraphics[width=\linewidth]{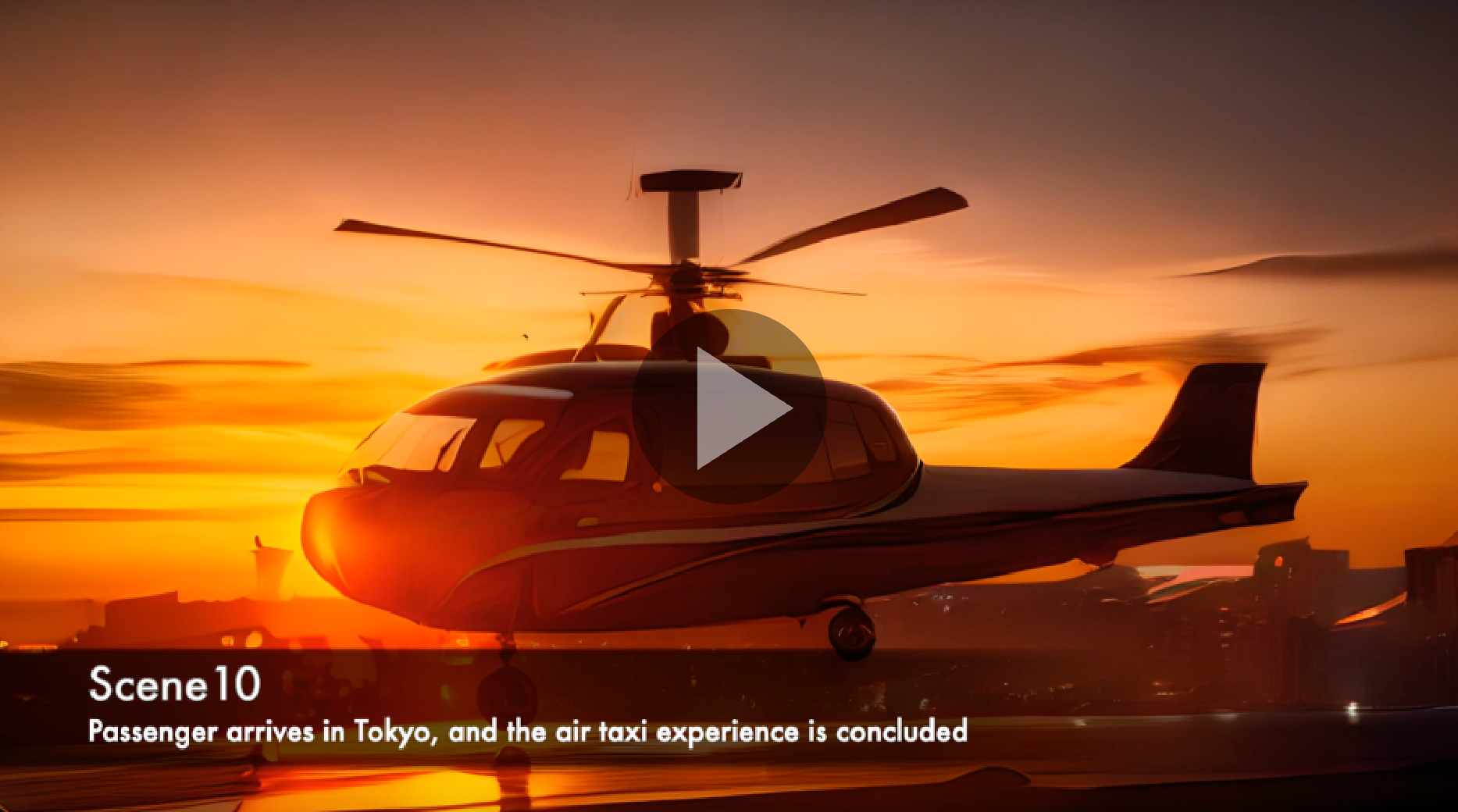}

  \label{fig:101}
\end{figure}
How satisfied are you with the services provided in the above simulation journey? \\
Strongly Unsatisfied,	Unsatisfied,	neutral,	Satisfied,	Strongly Satisfied\\
20. If this was your first experience with the air taxi service, would you consider using it again in the future? \\
Strongly impossible,	impossible,	neutral,	possible,	Strongly possible\\

\section{Prompt}

Based on the user-centred study we have designed empathize, define, ideate, and prototype prompts.

\begin{table}[H]
\centering
    \caption{The content of Prompt-ALOs}
    \label{tab:freq-31}
    \begin{tabular}{p{100mm}l}
        \toprule
        \multicolumn{1}{c}{Prompt 0: Prompt - ALOs}\\
        \midrule
        1. Create Abstract Language Objects (ALOs) for input using steps 1-11.\\ 
         2. Define mainObj with subObjList or Skip. Birth of ALOs affects all other ALOs.\\ 
          3. Add skills and knowledge to subObjList or Reload.\\ 
           4. Set specific states for subObjList or Reload.\\ 
            5. Validate initial state meets conditions or Skip.\\ 
           6. Update subObjList for state detection or Reload.\\ 
             7. Create managerObj with initial state or Reload.\\ 
             8. Update managerObj state using skills and knowledge. \\ 
             9. Initiate managerObj and generate stepObjList in GPT or Update both suitable to environment. \\ 
             10. Convert ALOs into GPT markdown scripts. Define object functions progressively, preserving features. \\ 
             11. Reference objects by name. Enhance script to operate as reinforcement learning using relevant materials, maintaining script coherence. \\ 
             12. Implement linguistic adjustments to prevent and rectify errors.\\ 
        \bottomrule
    \end{tabular}
\end{table}

\begin{table}[H]
\centering
    \caption{The content of Prompt-Empathize}
    \label{tab:freq-1}
    \begin{tabular}{p{100mm}l}
        \toprule
        \multicolumn{1}{c}{Prompt 1: Prompt-Empathize}\\
        \midrule
        0. Create Abstract Language Objects (ALOs) for {input} using steps 1-3.\\ 
        1. ALOs and brainstorm all parameters step-by-step to add and fill.\\
        2. get ALOs object and brainstorm to fill subobject parameters and output one ALOs object subobject list and parameters in table.\\
        3. Simulate ALOs based on all subobject list and parameters.\\
        4. Expand the simulation into a concrete storyboard based on all subobject list and parameters.\\
        5. Analyse the user pain point, motivation, user needs based on the storyboard in table.\\
        \bottomrule
    \end{tabular}
\end{table}

\begin{table}[H]
\centering
    \caption{The content of Prompt-Define}
    \label{tab:freq-2}
    \begin{tabular}{p{100mm}l}
        \toprule
        \multicolumn{1}{c}{Prompt 2: Prompt-Define}\\
        \midrule
        0. Create Abstract Language Objects (ALOs) for {input} using steps 1-3.\\ 
        1. generate a persona, which includes pain points, goals, and personality based on the form1:\\
        User Pain Points:\\
        - Long commute times\\
        - Traffic congestion\\
        - Limited privacy in traditional transportation\\
        - Environmental impact of transportation\\
        Motivation:\\
        - Save time during daily commute\\
        - Increase comfort and privacy\\
        - Reduce environmental impact\\
        User Needs:\\
        - Fast, reliable transportation option\\
        - Availability during peak travel times\\
        - Affordability within their budget\\
        - Eco-friendly transportation solution\\
        2. simulate scenario, describes how ALOs can best help persona achieve motivation, from persona‘s perspective\\
        3. generate user needs based on scenario in table\\
        \bottomrule
    \end{tabular}
\end{table}

\begin{table}[H]
\centering
    \caption{The content of Prompt-Ideate}
    \label{tab:freq-3}
    \begin{tabular}{p{100mm}l}
        \toprule
        \multicolumn{1}{c}{Prompt 3: Prompt-Ideate}\\
        \midrule
        0. Create Abstract Language Objects (ALOs) for {input} using steps 1-2.\\ 
        1. Brainstorm and come up with creative solution based on form2:\\
        \quad- Availability of air taxi service during peak hours\\
        \quad- Booking and payment options that are user-friendly and secure\\
        \quad- Real-time information on the air taxi's location and schedule\\
        \quad- Safe and comfortable air taxis with limited passenger capacity\\
        \quad- Efficient support service for any complaints or concerns.\\
        2. generate design requirement of ALOs based on creative solution\\
        \bottomrule
    \end{tabular}
\end{table}

\begin{table}[H]
\centering
    \caption{The content of Prompt-Prototype}
    \label{tab:freq-4}
    \begin{tabular}{p{100mm}l}
        \toprule
        \multicolumn{1}{c}{Prompt 4: Prompt-Prototype}\\
        \midrule
        0. Create Abstract Language Objects (ALOs) for {input} using steps 1-3.\\ 
        1. Create mockup of ALOs based on the following design requirements: \\
        \quad- Easy-to-use booking platform via a website or an application with reservation options and payment processing\\
        \quad- Integrating GPS technology for real-time tracking and status updates\\
        \quad- Air taxis designed to accommodate a limited number of passengers with comfortable seating and ample personal space\\
        \quad- An efficient customer support system to attend to user inquiries, complaints, or issues instantly\\
        \quad- Ensuring highly maintained and safe air taxis that strictly follow aviation and safety regulations\\
        2. Get mockup and brainstorm to fill subobject parameters and output one ALOs object subobject list and parameters\\
        3. generate concrete storyboard based on mockup with all subobject list and parameters to simulate ALOs interaction\\
        \bottomrule
    \end{tabular}
\end{table}

\section{Outputs from GPT-4}

\begin{table}[H]
\centering
    \caption{One of the results from Prompt-Empathize}
    \label{tab:1}
    \begin{tabular}{p{100mm}l}
        \toprule
        \multicolumn{1}{c}{The result of Prompt-Empathize}\\
        \midrule
        User Pain Points:\\
 \quad-Long commute times\\
 \quad-Traffic congestion\\
 \quad-Limited privacy in traditional transportation\\
 \quad-Environmental impact of transportation\\
 Motivation:\\
 \quad-Save time during daily commute\\
 \quad-Increase comfort and privacy\\
 \quad-Reduce environmental impact\\
User Needs:\\
 \quad-Fast, reliable transportation option\\
 \quad-Availability during peak travel times\\
 \quad-Affordability within their budget\\
 \quad-Eco-friendly transportation solution\\
        \bottomrule
    \end{tabular}
\end{table}

\begin{table}[H]
\centering
    \caption{One of the results from Prompt-Define}
    \label{tab:2}
    \begin{tabular}{p{100mm}l}
        \toprule
        \multicolumn{1}{c}{The result of Prompt-Define}\\
        \midrule
        - Availability of air taxi service during peak hours\\
- Booking and payment options that are user-friendly and secure\\
- Real-time information on the air taxi's location and schedule\\
- Safe and comfortable air taxis with limited passenger capacity\\
- Efficient support service for any complaints or concerns\\
        \bottomrule
    \end{tabular}
\end{table}

\begin{table}[H]
\centering
    \caption{One of the results from Prompt-Ideate}
    \label{tab:3}
    \begin{tabular}{p{100mm}l}
        \toprule
        \multicolumn{1}{c}{The result of Prompt-Ideate}\\
        \midrule
-Design an electric air taxi that is energy-efficient and environmentally friendly.\\
-Ensure that the air taxi has a sleek, modern design that appeals to passengers.\\
-Design a user-friendly mobile app and website that enable passengers to easily book, manage, and cancel air taxi rides.\\
-Include in-app functionality for passengers to provide feedback and rate their air taxi experience, helping the service continually improve and maintain high-quality standards.\\
        \bottomrule
    \end{tabular}
\end{table}

\begin{table}[H]
\centering
    \caption{One of the results from Prompt-Prototype}
    \label{tab:4}
    \begin{tabular}{p{100mm}l}
        \toprule
        \multicolumn{1}{c}{The Mockup result of Prompt-Prototype}\\
        \midrule
Booking Platform Parameters:\\
\quad-Website or application access\\
\quad-Reservation options\\
\quad-Payment processing\\
GPS Tracking Parameters:\\
\quad-Real-time location updates\\
\quad-Status updates\\
Air Taxi Design Parameters:\\
\quad-Limited passenger capacity\\
\quad-Comfortable seating\\
\quad-Ample personal space\\
Customer Support Parameters:\\
\quad-Instant response\\
\quad-Inquiry handling\\
\quad-Complaint resolution\\
\quad-Issue assistance\\
Safety Regulations Parameters:\\
\quad-Aviation regulations compliance\\
\quad-Regular maintenance\\
\quad-Safety checks\\
        \bottomrule
    \end{tabular}
\end{table}

\begin{table}[H]
\centering
    \caption{One of the storyboard results from Prompt-Prototype}
    \label{tab:5}
    \begin{tabular}{p{100mm}l}
        \toprule
        \multicolumn{1}{c}{The Mockup result of Prompt-Prototype}\\
        \midrule
Scene 1: A passenger (ALOs(passenger)) opens the air taxi booking platform (ALOs(air taxi)) on their smartphone.\\
Scene 2: The passenger browses the available air taxi trips and makes a reservation, securely processing their payment through the booking platform.\\
Scene 3: The passenger receives real-time GPS tracking updates on their air taxi's location and status.\\
Scene 4: Upon boarding the air taxi, the passenger finds comfortable seating and ample personal space, as per the air taxi design parameters.\\
Scene 5: During the trip, the passenger has a question and quickly receives assistance from the customer support team through the booking platform.\\
Scene 6: The air taxi adheres to all safety regulations, ensuring a secure and reliable trip for the passenger.\\
        \bottomrule
    \end{tabular}
\end{table}

\begin{table}[H]
\centering
    \caption{The mockup selected for user test}
    \label{tab:freq-5}
    \begin{tabular}{p{100mm}l}
        \toprule
        \multicolumn{1}{c}{Mockup}\\
        \midrule
        1. App-based Platform\\
        \quad- User Interface: Sleek design, intuitive navigation\\
        \quad- Payment: Credit Card, PayPal, Google Pay, Apple Pay\\
        \quad- Booking: Search air taxis, select time and location, confirm booking\\
        \quad- Profile: Name, contact information, payment options, booking history\\
        2. Real-time Tracking\\
        \quad- GPS: Google Maps integration\\
        \quad- Updates: Push notifications, app messages, arrival and departure time notifications\\
        3. Limited Seating Capacity Air Taxis\\
        \quad- Seating Arrangement: Ergonomic seats, 2-4 passengers capacity\\
        \quad- Safety Features: Seat belts, emergency exits, fire extinguishers\\
        \quad- Regulations: Compliant with aviation authorities and guidelines\\
        4. AI Customer Support\\
        \quad- Communication: Text-based chatbot, voice assistance\\
        \quad- Assistance: Respond to queries, handle complaints, assist with booking and cancellations\\
        \quad- Availability: In-app, website, email, phone\\
        \bottomrule
    \end{tabular}
\end{table}

\begin{table}[H]
\centering
    \caption{The storyboard selected for user test}
    \label{tab:freq-6}
    \begin{tabular}{p{100mm}l}
        \toprule
        \multicolumn{1}{c}{Storyboard}\\
        \midrule
        Scene1. Passenger opens the app and logs in to their profile.\\ 
        Scene2. Passenger navigates to the booking section and inputs their travel requirements: from Tsukuba to Tokyo.\\
        Scene3. The app shows a list of available air taxi with their respective times and prices.\\
        Scene4. Passenger selects their preferred air taxi and time and proceeds to the payment section.\\
        Scene5. Passenger confirms their payment through a secure payment method, like PayPal.\\
        Scene6. The booking is confirmed, and Passenger gets a confirmation message with details about the air taxi's departure time and location.\\
        Scene7. On the day of the departure, Passenger receives real-time updates on the air taxi's location and schedule through push notifications.\\
        Scene8. Passenger arrives at the departure point and boards the air taxi. They find a comfortable seating arrangement and safety features, such as seat belts and emergency equipment.\\
        Scene9. During the flight, Passenger has a question about the air taxi etiquette. They use the app to contact AI Customer Support, which promptly responds to their query.\\
        Scene10. Passenger arrives in Tokyo, and the air taxi experience is concluded.\\
        \bottomrule
    \end{tabular}
\end{table}

\section{Specific Data}

\begin{table}[H]

\caption{Results of a quantitative analysis of factors influencing Virtual flight to improve participants' attitudes towards the air taxi. This includes response from real people and virtual response from the GPT-4 and Gemini.}
\label{tab:freq-8}
\resizebox{\textwidth}{!}{
\begin{tabular}
{p{3.7cm}<{\centering} p{0.8cm}<{\centering} p{0.8cm}<{\centering} p{0.8cm}<{\centering} p{0.8cm}<{\centering} p{0.8cm}<{\centering} p{0.8cm}<{\centering} p{0.8cm}<{\centering} p{1cm}<{\centering} p{1cm}<{\centering}}\\
\hline
& N & \multicolumn{3}{c}{Mean} & \multicolumn{3}{c}{SD} & $F$&$ p $  \\ \hline
\multirow{3}{*}{}& \multirow{3}{*}{} & \multirow{3}{*}{Real} & \multirow{3}{*}{GPT-4}& \multirow{3}{*}{Gemini} & \multirow{3}{*}{Real} & \multirow{3}{*}{GPT-4} & \multirow{3}{*}{Gemini}&  Real& Real  \\ 
& & & && & & & GPT-4  & GPT-4             \\
& & & && & & & Gemini  & Gemini             \\
\hline
The attitude towards the air taxi (before virtual flight) & & & & & & & & $3.550$& $0.064$ \\
male & 40 & $3.60$ & $4.12$ & $4.07$ & $1.30$ & $0.76$ & $0.27$ & $15.671$  & $\textless{}0.001$* \\
female & 32 & $3.28$ & $3.25$ & $4.06$ & $1.02$& $1.05$ & $0.44$ &$0.043$ & $0.835$   \\
The attitude towards the air taxi (after virtual flight)  & & & & & & & & &  \\
male  & 40 & $4.07$ & $4.25$ & $4.42$& $0.80$ & $0.74$&$ 0.50$&&  \\
female & 32& $3.66$ & $3.41$& $4.47$& $0.70$& $1.19$& $0.51$& &   \\ 
\hline
The attitude towards the air taxi (before virtual flight) &&&&& & & & $3.733$& $0.008$*    \\
High school& 9& $2.22$& $3.67$& $4.00$& $1.09$& $0.87$ & $0.00$ & $1.688$  & $0.163$ \\
College & 15 & $2.93$ & $3.27 $& $4.07$ & $1.10$ & $1.10 $ & $0.26$& $0.162$&  $0.957$ \\
Bachelor & 28 & $3.68$ & $3.71$ & $4.04$ &$1.02$& $0.98$ & $0.33$& & \\
Master & 14  & $4.07$ & $4.14$ &$ 4.21$ & $1.07$ & $0.95$ & $0.43$ & &     \\
Ph.d.~& 6 &$4.17$ & $4.17$ & $4.00$ & $0.75$ & $0.75$ & $0.63 $ &  &    \\
The attitude towards the air taxi (after virtual flight)  &&&&&&&&&    \\
High school & 9&$ 4.00$& $4.00$ &$ 4.56$& $1.12$& $1.00$&$ 0.53$ &  &       \\
College& 15& $3.67$& $3.33$ &$ 4.47$& $0.72 $&$1.17$ &$ 0.52$ &  &  \\
Bachelor  & 28 & $3.79$& $3.89$& $4.46 $& $0.69 $& $1.03$& $0.51$&    &   \\
Master   & $14$ & $4.14$& $4.21$& $4.36$& $0.77$ &$0.97 $& 0.50 &  & \\
Ph.d.~& 6 & $4.17$ & $4.17$& $4.33$ & $0.75$ & $0.75$  &$0.52$& &  \\ \hline
The attitude towards the air taxi (before virtual flight) & &&&&&& & $0.124$  & $0.726$   \\
18$\sim$40 & 51 & $3.45$ & $3.98$ & $4.04$ & $1.12$ & $0.84$ & $0.20$& $0.222$ & $0.825$ \\
40+ & 21& $3.48$& $3.14$& $4.14$& $1.36$ & $1.11$ & $0.57$ & $12.964$& $\textless{}0.001$* \\
The attitude towards the air taxi (after virtual flight)  &&&&&&&&&\\
18$\sim$40  & 51 &$3.94$ &$4.14$ & $4.51$ & $0.78$& $0.87$& $0.51$ & & \\
40+  & 21 & $3.76$&$3.24$&$4.29 $& $0.77$& $1.18$ & $0.46$ & & \\ \hline
The attitude towards the air taxi (before virtual flight) & & & & & & & & $0.201$& $0.655$\\
unemployed & 9 &$ 3.44 $&$2.56$& $3.89$ & $0.73$& $0.88$& $0.33$& $20.195$ & $\textless{}0.001$* \\
employed & 63 & $3.46 $& $3.91$& $4.09$&$1.24$ &$ 0.89$ &$ 0.35$ & $2.259$&$0.137$\\
The attitude towards the air taxi (after virtual flight) &&&&&&&&&   \\
unemployed   & 9 &$ 3.67$ & $2.56$& $4.33$& $0.71$&$0.88$&$0.50$ &&  \\
employed & 63&$3.92$ & $4.06$ & $4.46$& $0.79$& $0.93$ & $0.50$ && \\ \hline
\end{tabular}
}
\begin{tablenotes}
    \footnotesize
    \item Real: Real Participants
    \item GPT-4: GPT-4 Participants
    \item Gemini: Gemini Participants
    \item *$p\textless{}.05$; **$p\textless{}.01$; ***$p\textless{}.001$
    \end{tablenotes}
\end{table}

\begin{table}[H]

\caption{The results of a quantitative analysis of influencing factors affecting Real, GPT-4, and Gemini responses' satisfaction with the flight simulation journey.*$p\textless.05$; **$p\textless.01$; ***$p\textless.001.$}
\label{tab:freq-20}
\resizebox{\textwidth}{!}{
\begin{tabular}
{p{3.4cm}<{\centering} p{0.3cm}<{\centering} p{0.9cm}<{\centering} p{0.9cm}<{\centering} p{1.0cm}<{\centering} p{0.7cm}<{\centering} p{0.7cm}<{\centering} p{0.9cm}<{\centering} p{1.3cm}<{\centering} p{1.2cm}<{\centering}}
\hline
& N & \multicolumn{3}{c}{Mean} & \multicolumn{3}{c}{SD} & $t$/$F$& $p$   \\ \hline
\multirow{3}{*}{}& \multirow{3}{*}{} & \multirow{3}{*}{Real} & \multirow{3}{*}{GPT-4}& \multirow{3}{*}{Gemini} & \multirow{3}{*}{Real} & \multirow{3}{*}{GPT-4} & \multirow{3}{*}{Gemini}&  Real& Real  \\ 
& & & && & & & GPT-4  & GPT-4             \\
& & & && & & & Gemini  & Gemini             \\
\hline
The satisfaction score toward Air Taxi Journey& & & & & & & & 2.508& 0.014* \\
male & 40 & 3.97& 4.17 & 4.02 & 0.83 & 0.75& 0.42 & 3.286 & 0.002* \\
female & 32 & 3.53 & 3.50 & 4.09 & 0.62& 0.95 & 0.30 &-0.779 & 0.439  \\
\hline
The satisfaction score toward Air Taxi Journey&&&&& & & & &   \\
High school& 9& 3.78&3.56& 3.89& 0.97& 0.88 &0.60 &1.319& 0.272 \\
College & 15 & 3.60 &3.33 & 4.00 & 0.74 & 0.98  &0.00& 3.543& 0.011* \\
Bachelor & 28 & 3.64 & 3.93 & 4.07 &0.62&0.86 & 0.38&0.162 & 0.957\\
Master & 14  & 4.07 &4.29 & 4.21 & 0.92 & 0.73 & 0.43 & &     \\
Ph.d.~& 6 &4.17 & 4.50& 4.00 & 0.75 & 0.56 & 0.00  &  &    \\
\hline
The satisfaction score toward Air Taxi Journey&&&&& & & & -0.222& 0.825 \\
18$\sim$40 & 51 & 3.76 & 4.04 & 4.04 & 0.76 & 0.77 & 0.40& 2.174  & 0.038*   \\
40+ & 21& 3.81& 3.48& 4.09& 0.81 & 1.08 & 0.30& -0.580& 0.564 \\
\hline
The satisfaction score toward Air Taxi Journey&&&&& & & & 0.000& 1.000 \\
unemployed & 9 & 3.78 &2.89& 4.05 & 0.67& 0.78& 0.38& -3.824& $\textless{}0.001$* \\
employed & 63 &3.78 & 4.02& 4.11&0.79 & 0.83 &0.33 & -0.478& 0.634\\
\hline
\end{tabular}
}
\begin{tablenotes}
    \footnotesize
    \item Real: Real Participants
    \item GPT-4: GPT-4 Participants
    \item Gemini: Gemini Participants
    \item *$p\textless{}.05$; **$p\textless{}.01$; ***$p\textless{}.001$
    \end{tablenotes}
\end{table}

\end{document}